\documentclass[11pt]{article}
\usepackage{abstract}
\usepackage{amsbsy}
\usepackage{amscd}
\usepackage{amsfonts}
\usepackage{amsmath}
\usepackage{amssymb}
\usepackage{amsthm}
\usepackage{bm}
\usepackage{calligra}
\usepackage{cite}
\usepackage{epsfig}
\usepackage{epstopdf}
\usepackage{float}
\usepackage{graphicx}
\usepackage{overpic}
\usepackage[margin=0.5in]{geometry}
\usepackage{mathrsfs}
\usepackage{mathtools}
\usepackage{multirow}
\usepackage{xcolor}
\usepackage[T1]{fontenc}
\usepackage{lmodern}
\usepackage{psfrag}
\usepackage{subfigure}
\usepackage{booktabs}
\usepackage{listings}
\usepackage{longtable}
\usepackage{latexsym}
\usepackage{arydshln}
\usepackage{enumerate}
\usepackage{verbatim}
\usepackage{caption}
\allowdisplaybreaks[4]

\usepackage{changes}
\usepackage{booktabs}

\date{}
\title{Orbital dynamics and gravitational wave of Schwarzschild-de Sitter black hole immersed in perfect fluid dark matter}

\author{Ruifang~Wang\textsuperscript{1,2,}\thanks{\href{mailto:wangruifang16@sina.com}{wangruifang16@sina.com}} , Fabao~Gao\textsuperscript{1,}\thanks{\href{mailto:gaofabao@sina.com}{gaofabao@sina.com}} , Ruihong~Yue\textsuperscript{2,}\thanks{\href{mailto:rhyue@yzu.edu.cn}{rhyue@yzu.edu.cn}}\\
\parbox{\textwidth}{\itshape\footnotesize
    \textsuperscript{1}School of Mathematics, Yangzhou University, Yangzhou 225009, P.R.\,China\\
    \textsuperscript{2}Center for Gravitation and Cosmology, College of Physical Science and Technology, Yangzhou University, Yangzhou 225009, P.R.\,China}}

\numberwithin{equation}{section}

\makeatletter
\renewcommand{\@fnsymbol}[1]{\@alph{#1}}
\makeatother

\usepackage[colorlinks,
            CJKbookmarks=true,
            linkcolor=blue,
            anchorcolor=red,
             urlcolor=blue,
            citecolor=blue
            ]{hyperref}

\begin{document}

\maketitle

\begin{abstract}
The dynamical properties of bound particle orbits around a Schwarzschild-de Sitter black hole immersed in perfect fluid dark matter (PFDM) are investigated. An analysis of the horizon topology shows that both event and cosmological horizons emerge within specific parametric regimes. Curvature invariants, including the Ricci scalar, Ricci-squared, and Kretschmann scalar, confirm the existence of a curvature singularity at $r = 0$. By utilizing the pericenter precession data of the S2 star in Sgr A*, the PFDM parameter is constrained to $0 < \lambda < 7.726412 \times 10^{5}$ M$_{\odot}$. For a source moving along a periodic orbit, the emitted gravitational waveforms exhibit two key features: (i) the orbital period increases with the ``zoom'' number, accompanied by a modest growth in wave amplitude, and (ii) the PFDM parameter nonlinearly modulates the period, leading to a slight amplitude suppression and introducing only a negligible phase shift relative to the vacuum case. These results explicitly characterize the local imprints of PFDM on relativistic orbital dynamics and offer predictions that are testable by future high-precision astrometric and gravitational-wave observations.

\vskip 0.2cm
\noindent {\bf Key words:} {Black hole; dark matter; periodic orbit; gravitational wave}
\end{abstract}

\section{Introduction}\label{se-1}
General Relativity (GR), since its formulation, has firmly established itself as a cornerstone of modern physics. Black holes (BHs), as extreme astrophysical entities predicted by GR, represent the most powerful gravitational sources in the universe and are generally expected to possess high spin and strong magnetic fields. These characteristics make them ideal natural laboratories for investigating the behavior of matter and the properties of gravity under extreme conditions. In recent years, multiple lines of observational evidence have substantiated their existence. A major breakthrough came with the detection of gravitational waves from binary black hole mergers by the LIGO-Virgo collaborations \cite{Abbott2016a, Abbott2016b}, followed by the direct imaging of the shadow of the M87* black hole \cite{Akiyama2019} and the subsequent release of the image of Sgr A* by the Event Horizon Telescope \cite{Akiyama2022} using very-long-baseline interferometry. Moreover, electromagnetic observations of accretion disk spectra also provide supporting evidence for black holes \cite{Frank2002, Yuan2014}. These landmark achievements not only open new pathways for probing and experimentally verifying GR in regimes of strong gravity, but also substantially enhance our understanding of accretion physics in the strong-gravity regime near event horizons. Crucially, they deliver compelling empirical evidence affirming the validity of GR under extreme gravitational fields, while simultaneously offering valuable avenues to test alternative theories of gravity.

According to the Standard Cosmological Model supported by indirect observations, the universe's energy-mass budget is dominated by dark energy ($\sim 68\%$) and dark matter ($\sim 27\%$), while baryonic matter constitutes only about 5\% \cite{Ade2014}. Cosmologically, dark matter plays a fundamental gravitational role in clustering baryonic matter, thereby enabling the formation of galaxies, clusters, and supercluster-scale structures. However, its non-interacting nature with electromagnetic radiation precludes any emission, absorption, or reflection of light, rendering dark matter undetectable through direct electromagnetic means. Recently, gravitational waves emitted by perturbed black holes have recently emerged as a promising indirect avenue for probing dark matter \cite{Zhang2025a, Zhang2025b}.

Investigating black hole behavior in these astrophysical environments composed of different dark matter candidates is essential, since dark matter influences geodesic motion and consequently alters both dynamical and energetic properties of the system. In response, a variety of theoretical frameworks have been developed to investigate dark matter-black hole interactions. A widely adopted approach models the dark matter halo with astrophysical density profiles parameterized by central density and scale radius, wherein dark matter forms halo structures around black holes and contributes to their effective mass \cite{Navarro1996, Kiselev2003, Kiselev2005, Jusufi2019, Tanaka2025}. In such models, the dark matter density reaches its peak near the black hole and declines asymptotically at large distances. Several well-known density profiles commonly applied in galactic structure simulations within the cold dark matter scenario belong to this category \cite{Navarro1997, Burkert1995, Xu2018a, Wang2020, Xu2020}. An alternative description treats dark matter as a perfect fluid \cite{Kiselev, Rahaman2011, Li2012b}, which provides a well-defined stress-energy tensor that frequently admits analytical solutions. In contrast to conventional particle-based descriptions, the PFDM model represents dark matter as a continuous, inviscid medium whose dynamical behavior is governed by prescribed equations of state. This formalism facilitates the study of gravitational interactions between dark matter and black holes, including accretion processes and modifications to spacetime geometry. An extended model where a PFDM-black hole system coexists with an inhomogeneous phantom field was developed by Li et al. \cite{Li2012b}. Subsequently, rotating black hole solutions in Kerr-type spacetimes surrounded by PFDM were constructed by Xu et al. \cite{Xu2018b} using the Newman-Janis algorithm. This framework effectively extends the known Kerr-de Sitter/anti-de Sitter framework with a cosmological constant. In recent years, the PFDM model has been extensively applied to investigate diverse black hole configurations, including metric modifications of Schwarzschild, Kerr, Bardeen and Euler-Heisenberg solutions \cite{Hamil2025a, Rizwan2019, Zhang2021, Ma2024}, geodesic motion \cite{Das2021, Rayimbaev2021, Das2024}, the implications for black hole shadows \cite{Hou2018, Haroon2019, Hamil2025b}, thermodynamic properties \cite{Liang2023a, Rakhimova2023, Rizwan2025}, gravitational lensing \cite{Atamurotov2021, Qiao2023}, quasinormal modes \cite{Jusufi2020, Chen2024}.

Geodesic analysis serves as a fundamental tool for probing the gravitational dynamics of black hole spacetimes, providing critical insights into both geometric properties and causal structure through the study of particle trajectories. In black hole spacetimes, null geodesics define key observational features including photon spheres, shadow boundaries, and gravitational lensing patterns, whereas timelike geodesics dictate the orbital motion and relativistic precession of massive objects. The examination of geodesic equations further elucidates fundamental phenomena such as perihelion precession, gravitational time delay, and light deflection, making geodesic analysis an increasingly vital component in contemporary black hole research \cite{Cardoso2009, Ianniccari2024, Al-Badawi2025}. 

Motivated by these results, we investigate in detail the dynamic characteristics of the orbit and the periodic orbits' gravitational wave around the Schwarzschild-de Sitter (dS) BH immersed in PFDM. The rest of this paper is structured as follows. Section \ref{se-2} presents the BH metric, the horizon structure, and three curvature invariants. Section \ref{se-3} derives the timelike geodesic equations via the Euler-Lagrange equation and characterizes the general properties of bound orbits. Section \ref{se-4} employs the orbital classification method and the observation data of the S2 star orbiting Sgr A* to impose constraints on PFDM parameters, and also examines periodic orbits and their gravitational wave emission. Section \ref{se-5} concludes the paper with a summary and discussion.

\section{Metric, horizons, and scalar invariants}\label{se-2}
The action of the Schwarzschild-dS BH immersed in PFDM is given by \cite{Li2012a, Rayimbaev2021}
\begin{align}
\mathcal{I}=\frac{1}{16\pi}\int{{\rm d}^4x \sqrt{-g}\Big[R-2\Lambda-2\big(\nabla_\mu \phi \nabla^\mu \phi-2V\left(\phi\right)\big)-4\left(L_{DM}+L_I\right)\Big]},\label{1} 
\end{align}
with the Ricci scalar $R$, the cosmological constant $\Lambda$, the phantom field potential $V\left(\phi\right)$, the dark matter Lagrangian density $L_{DM}$, and $L_I$ being the coupling quantity of the dark matter and phantom field. By variating and calculating the action (\ref{1}), it leads to the Einstein field equation as follows
\begin{align}
R_{\mu\nu}-\frac{1}{2}g_{\mu\nu}R+\Lambda g_{\mu\nu}=8\pi\left(T_{\mu\nu}^{DM}+2\nabla_\mu \phi\nabla_\nu \phi-g_{\mu\nu}\nabla_\alpha \phi\nabla_\alpha \phi+T_{\mu\nu}^I\right),
\end{align}
where $T_{\mu\nu}^{DM}$ represents the energy-momentum tensor of the PFDM. The above equation admits an exact solution corresponding to the Schwarzschild-dS BH immersed in PFDM, with the metric given by
\begin{align}
{\rm d} s^2&=g_{\mu\nu}{\rm d}x^\mu{\rm d}x^\nu\nonumber\\
&=-f\left(r\right){\rm d} t^2+f^{-1}\left(r\right){\rm d}r^2+r^2{\rm d}\Omega^2,\label{2} 
\end{align}
where the lapse function and the metric on 2-dimensional unit sphere are
\begin{align}
f\left(r\right)=1-\frac{2M}{r}-\frac{\Lambda}{3}r^2+\frac{\lambda}{r}\ln\frac{r}{|\lambda|},\label{3}
\end{align}
and
\begin{align}
{\rm d}\Omega^2={\rm d}\theta^2+\sin^2{\theta}{\rm d}\phi^2,\label{4}
\end{align}
respectively. Here $M$ is the mass of black hole, and $\lambda$ is the parameter related to the dark matter field. Take $\lambda=0$, the metric (\ref{2}) describes the Schwarzschild-dS spacetime. For $\Lambda=\lambda=0$, the metric reduces to the Schwarzschild one. Note that in equation (\ref{3}), there contains an absolute value term of the parameter $\lambda$. Given that the density, radial pressure, and tangential pressure of the dark matter field distribution are all positive:
\begin{align}
\rho=-P_{r}=\frac{\lambda}{8\pi r^3}>0, \quad P_{\theta}=P_{\phi}=\frac{\lambda}{16\pi r^3}>0,\label{5}
\end{align}
a positive value of $\lambda$ is adopted in the following study, i.e., $\lambda>0$. 

\begin{figure}[H]
\centering
\subfigure[]{
\centering{\includegraphics[scale=0.38]{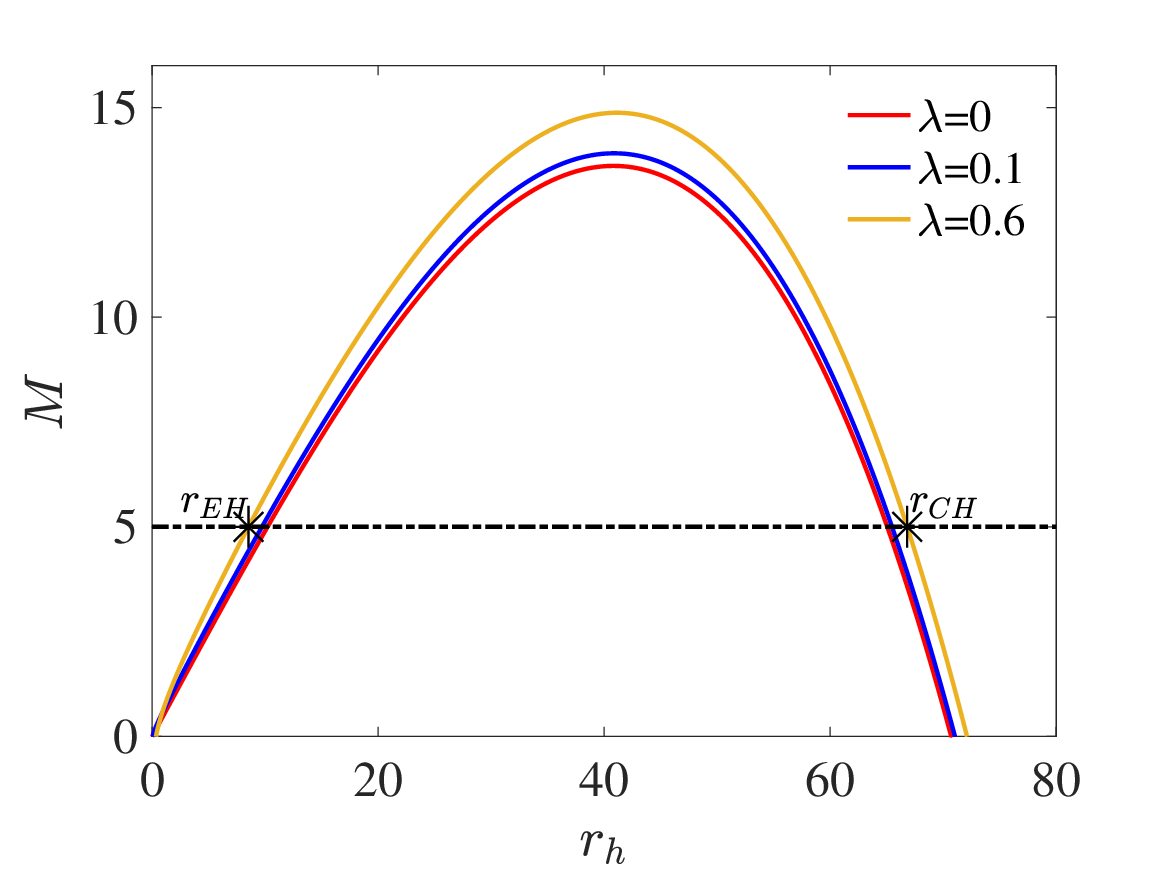}}\label{fig1a}}
\subfigure[]{
\centering{\includegraphics[scale=0.38]{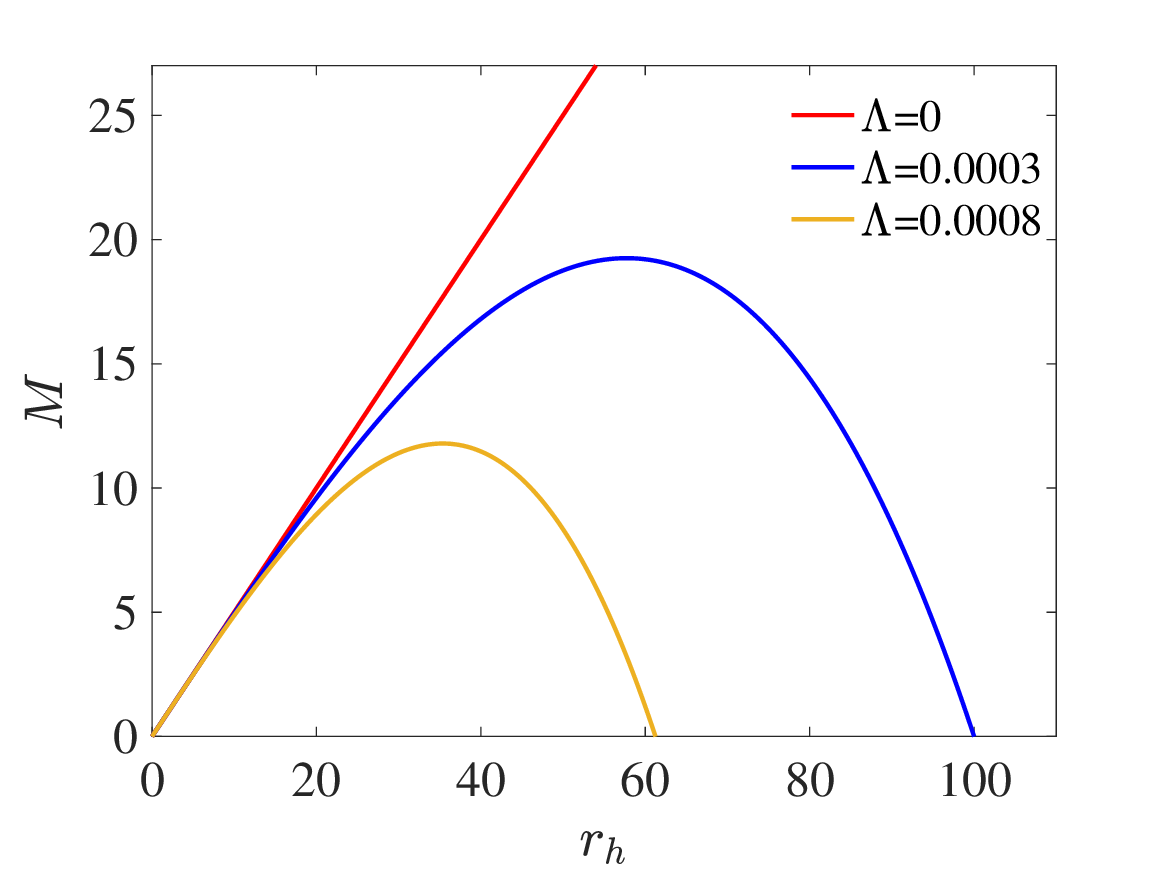}}\label{fig1b}}
\caption{Variation of the black hole's mass with its horizon radius for various values of parameters $\lambda$ and $\Lambda$.}
\label{fig1}
\end{figure}

\begin{figure}[H]
\centering
\subfigure[]{
\centering{\includegraphics[scale=0.38]{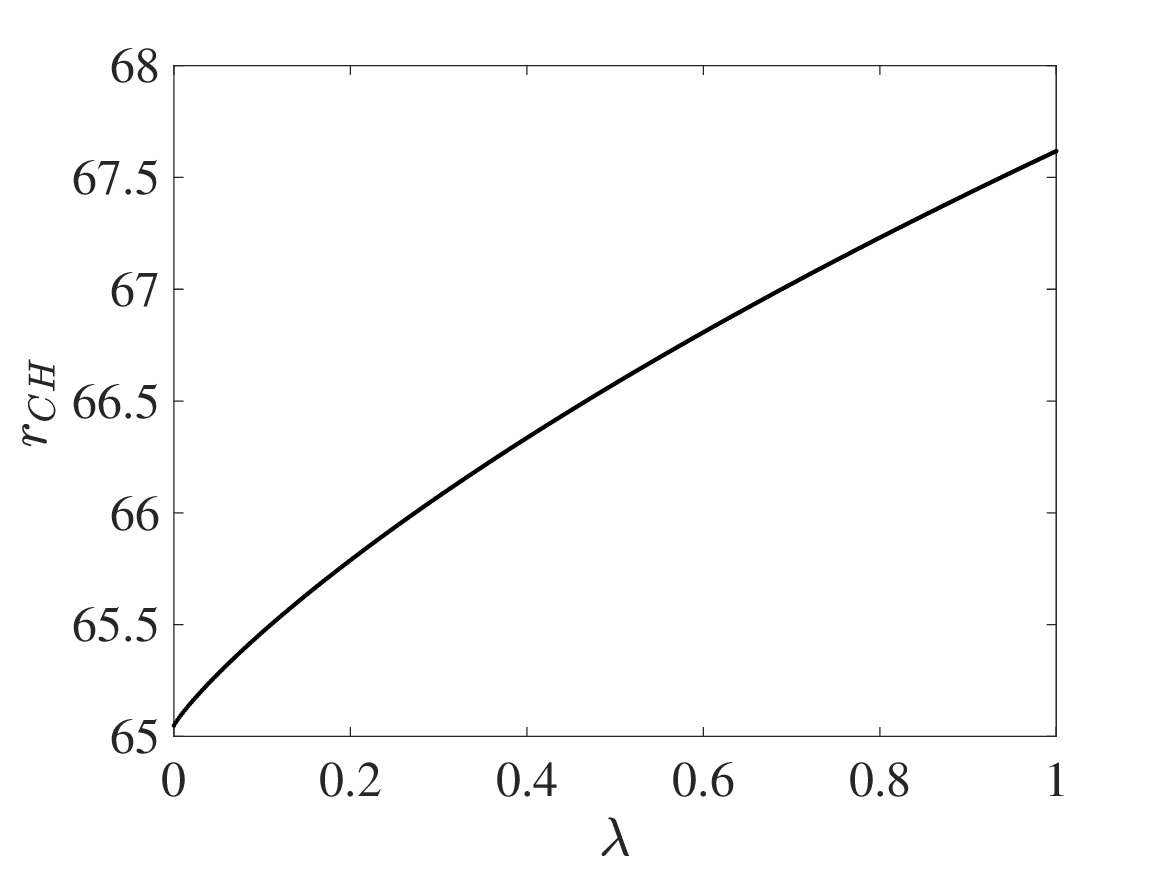}}\label{fig2a}}
\subfigure[]{
\centering{\includegraphics[scale=0.38]{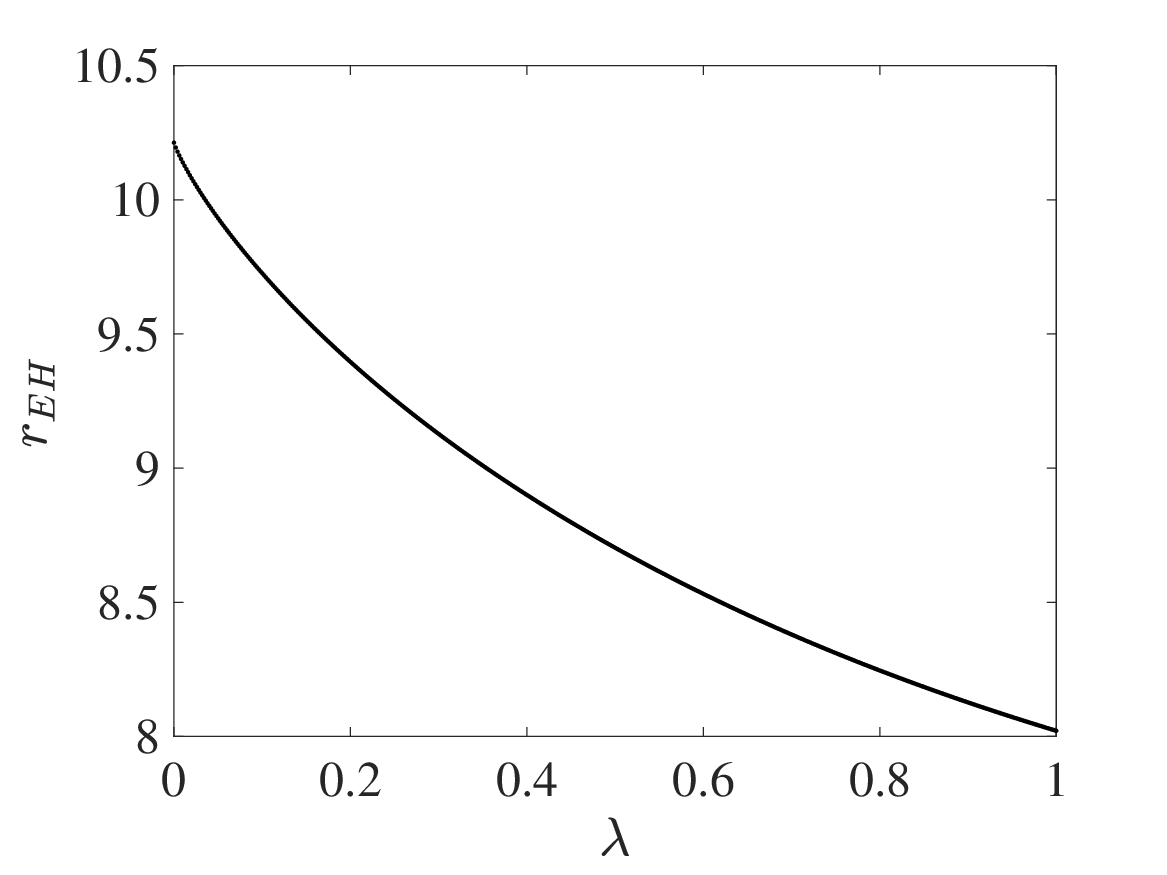}}\label{fig2b}}
\caption{Variation of the cosmological horizon and the event horizon with the parameter $\lambda$.}
\label{fig2}
\end{figure}

Focus on the lapse function (\ref{3}) and omit the absolute value symbol, the horizon $r_{h}$ of black hole is determined by
\begin{align}
f\left(r_{h}\right)=1-\frac{2M}{r_{h}}-\frac{\Lambda}{3}r_{h}^2+\frac{\lambda}{r_{h}}\ln\frac{r_{h}}{\lambda}=0.\label{6}
\end{align}
Rearrange above equation yields that the horizon determines the mass of black hole as follows
\begin{align}
M=\frac{1}{2}\left(r_{h}-\frac{\Lambda}{3}r_{h}^3+\lambda\ln\frac{r_{h}}{\lambda}\right).\label{7}
\end{align}
Since the above equation contains a logarithmic function, it is rather difficult to directly solve its analytical solution. Therefore, the existence of the event horizon of a black hole is examined using numerical methods below. By selecting appropriate parameter values, the desired results can be obtained. Accordingly, variation curves of the black hole mass with respect to the horizon for various values of $\Lambda$ and $\lambda$ are shown in Figure \ref{fig1}. This black hole may have two event horizons, one event horizon, or no event horizon at all. To be precise, choose $\Lambda=0.0006$ and different values of $\lambda$, the relation of $M$ and $r_{h}$ is shown in Figure \ref{fig1a}. It is obviously that given a suitable black hole mass, for example given $M=5$ (the black dotted line), there are two horizons, the larger one being the cosmological horizon $r_{CH}$ and the smaller one being the event horizon $r_{EH}$. As the parameter $\lambda$ value increases, the cosmological horizon increases, while the event horizon decreases, see Figure \ref{fig2} in detail. That is to say, the cosmological horizon (the event horizon) of the Schwarzschild-dS BH with PFDM is larger (smaller) than that without PFDM. In addition, a black hole has a maximum mass, which locates at the maximum point of the $r_{h}-M$ curve, and in this case the two horizons merge into one. Furthermore, when the mass of the black hole exceeds this maximum value, its horizons cease to exist.

Similarly, given the value of $\lambda$, the variation curves of mass with the cosmological constant can be seen in Figure \ref{fig1b}. Note that for $\Lambda=0$, the curve of $r_{h}-M$ is monotonous, which means that there exist only one horizon for a black hole of arbitrary mass, and the more greater the mass, the larger the horizon. While for $\Lambda\not=0$, the same variation occurs as the curves in Figure \ref{fig1a}. There may be one, two, or no horizons. Additionally, the cosmological horizon (the event horizon) of black hole decreases (increases) as the cosmological constant.

\begin{figure}[H]
\centering{\includegraphics[scale=0.38]{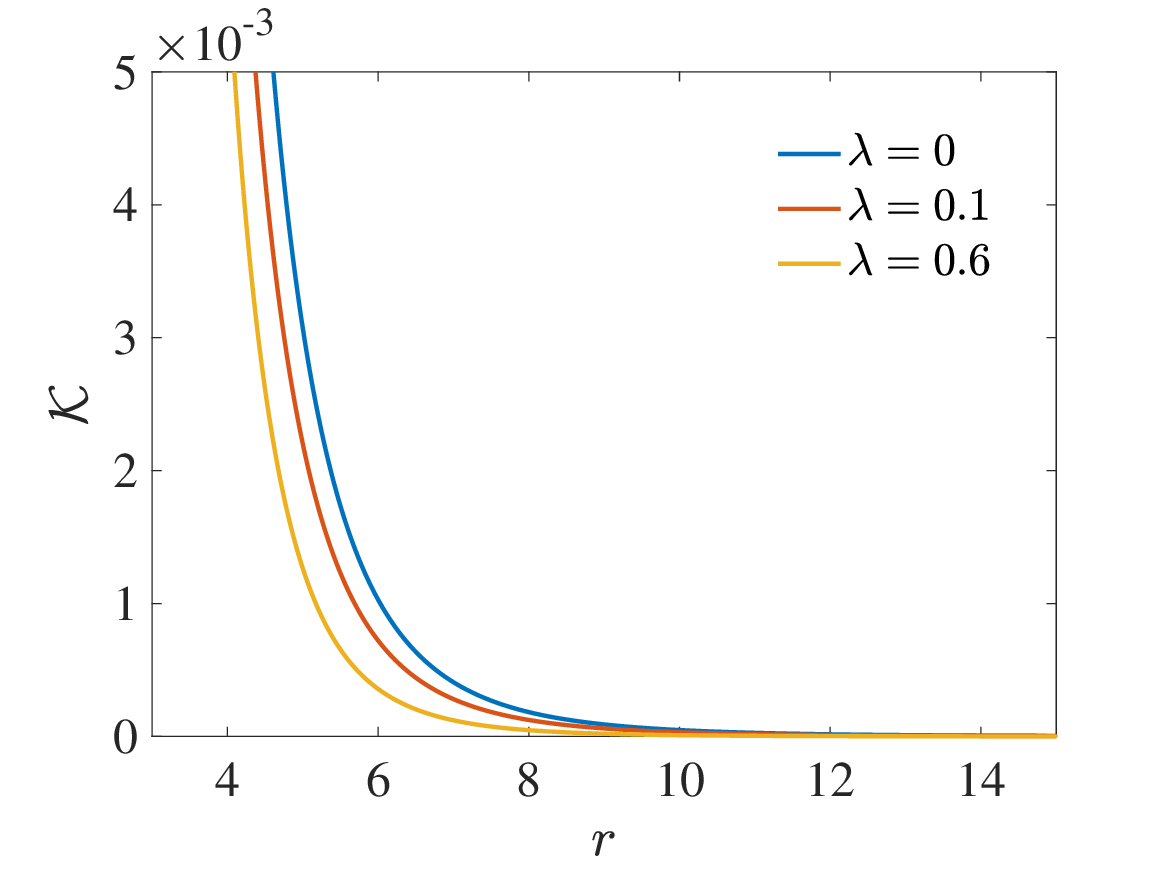}}
\caption{The curve of the Kretschmann scalar varying with the radial radius for various values of the parameter $\lambda$.}
\label{fig3}
\end{figure}

As is well known, in GR, the gravitational effects of black holes are described by the geometry of spacetime, represented by the metric. The curvature of a Schwarzschild black hole is characterized by the Kretschmann scalar ($\mathcal{K}= 48 M^2/r^6$), a geometric curvature invariant that diverges at the singularity. To study the geometric properties of the Schwarzschild-dS black hole immersed in PFDM, the complete analytical expressions for three scalar invariants of the metric (\ref{2}) are derived: the Ricci scalar $R$, the square of the Ricci tensor $R^{2}$, and the Kretschmann scalar $\mathcal{K}$, as follows.
\begin{align}
R&=g^{\mu\nu}R_{\mu\nu}=4 \Lambda -\frac{\lambda }{r^3}, \quad\quad R^{2}=R_{\mu\nu}R^{\mu\nu}=4 \Lambda ^2+\frac{5 \lambda ^2}{2 r^6}-\frac{2 \lambda  \Lambda }{r^3},\\
\mathcal{K}&=R_{\mu\nu\sigma\rho}R^{\mu\nu\sigma\rho}\nonumber\\
&=\frac{48 M^2}{r^6}+\frac{1}{3 r^6}\left(120 M \lambda+39 \lambda ^2-4 \lambda  \Lambda  r^3+8 \Lambda ^2 r^6-12 \lambda \left(12 M+5 \lambda \right) \ln \frac{r}{\lambda }+36 \lambda ^2 \ln ^2 \frac{r}{\lambda }\right).
\end{align}
The values of these scalar invariants are independent of the observer’s coordinate system and provide a true measure of the intrinsic curvature of spacetime. The calculation results indicate that:
\begin{itemize}
\item The presence of the cosmological constant and PFDM results in the Ricci scalar and the squares of the Ricci tensor being non-zero, indicating that both serve as significant ``source terms'' the curvature of spacetime.
\item Considering the cosmological constant and PFDM, as the radial distance $r$ approaches $0$, the Kretschmann scalar $\mathcal{K}$ still diverges, indicating the presence of an irremovable, geometrically significant physical singularity (curvature singularity) at $r = 0$, rather than a coordinate singularity.
\item Notably, larger values of the cosmological constant lead to the formation of the black hole nearer to the coordinate origin (details in Figure \ref{fig3}).
\end{itemize}

\section{Geodesics around the Schwarzschild-dS BH immersed in PFDM}\label{se-3}
With the metric of the Schwarzschild-dS BH immersed in PFDM established, attention turns to the motion of timelike particles along geodesics. Although spacetime curvature cannot be observed directly, the orbits of particles in gravitational fields offer a means to verify the predictions of GR.

\subsection{Equations of geodesic motion by Lagrangian function}
In order to investigate the geodesics around the Schwarzschild-dS BH immersed in PFDM, the motion of test particles with mass $m$ along the geodesics is taken into consideration by utilizing the Lagrangian function, which reads
\begin{align}
-\delta \equiv& \; 2\mathscr{L}=g_{\mu\nu} \dot{x}^\mu \dot{x}^\nu \nonumber\\
=&-\left(1-\frac{2M}{r}-\frac{\Lambda}{3}r^2+\frac{\lambda}{r}\ln\frac{r}{\lambda}\right)\dot{t}^2+\left(1-\frac{2M}{r}-\frac{\Lambda}{3}r^2+\frac{\lambda}{r}\ln\frac{r}{\lambda}\right)^{-1}\dot{r}^2+r^2\dot{\theta}^2+r^2\sin^2\theta\dot{\phi}^2,\label{8}
\end{align}
wherein $\delta=1$ for the timelike particle and $\delta=0$ for the photon, and $\dot{x}^\mu$ denotes the differentiation of $x^\mu$ with respect to the affine parameter $\tau$. Applying the Euler-Lagrange equation
\begin{align}
\frac{\rm d}{{\rm d}\tau}\left(\frac{\partial \mathscr{L}}{\partial \dot x^\mu}\right)-\frac{\partial \mathscr{L}}{\partial x^\mu}=0,\label{9}
\end{align}
it yields
\begin{align}
p_t&=\frac{\partial \mathscr{L}}{\partial \dot t}=-\left(1-\frac{2M}{r}-\frac{\Lambda}{3}r^2+\frac{\lambda}{r}\ln\frac{r}{\lambda}\right)\dot{t} \equiv -E,\label{10}\\
p_r&=\frac{\partial \mathscr{L}}{\partial \dot r}=\left(1-\frac{2M}{r}-\frac{\Lambda}{3}r^2+\frac{\lambda}{r}\ln\frac{r}{\lambda}\right)^{-1}\dot{r},\label{11}\\
p_\theta&=\frac{\partial \mathscr{L}}{\partial \dot \theta}=r^2\dot{\theta},\label{12}\\
p_\phi&=\frac{\partial \mathscr{L}}{\partial \dot \phi}=r^2 \sin^2\theta\dot{\phi} \equiv L,\label{13}
\end{align}
where $E$ and $L$ are the energy per unit mass and the angular momentum per unit mass of the particles, respectively. Consider the geodesic motion on the equatorial plane $\theta=\pi/2$. The scope of this paper is intentionally limited to the study of timelike geodesics; a systematic investigation of null geodesics remains a subject for future research. Thus, let $\delta=1$, and substitute equations (\ref{10})-(\ref{13}) into the Lagrangian (\ref{8}) to obtain the equation of radial motion of test particles as follows
\begin{align}
\dot{r}^2=E^2-\left(1-\frac{2M}{r}-\frac{\Lambda}{3}r^2+\frac{\lambda}{r}\ln\frac{r}{\lambda}\right)\left(1+\frac{L^2}{r^2}\right)\equiv E^2-V_{eff}\left(r\right),\label{14}
\end{align}
where $V_{eff}\left(r\right)$ denotes the effective potential of timelike particles.

\subsection{Bound timelike orbits}
\begin{figure}[H]
\centering
\subfigure[$\Lambda=0.0001,\; \lambda=0.01$]{
\centering{\includegraphics[scale=0.35]{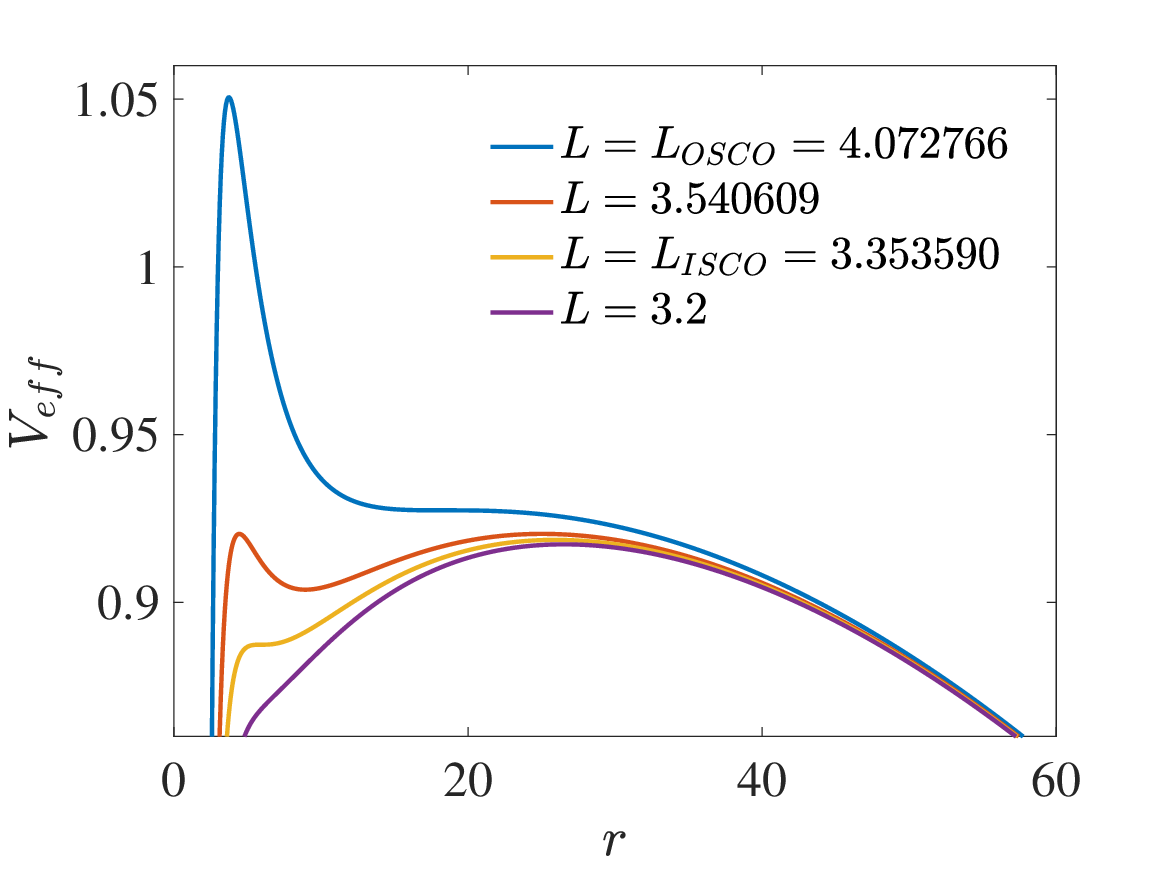}}\label{fig4a}}
\subfigure[$\Lambda=0.0001,\; \lambda=0.1$]{
\centering{\includegraphics[scale=0.35]{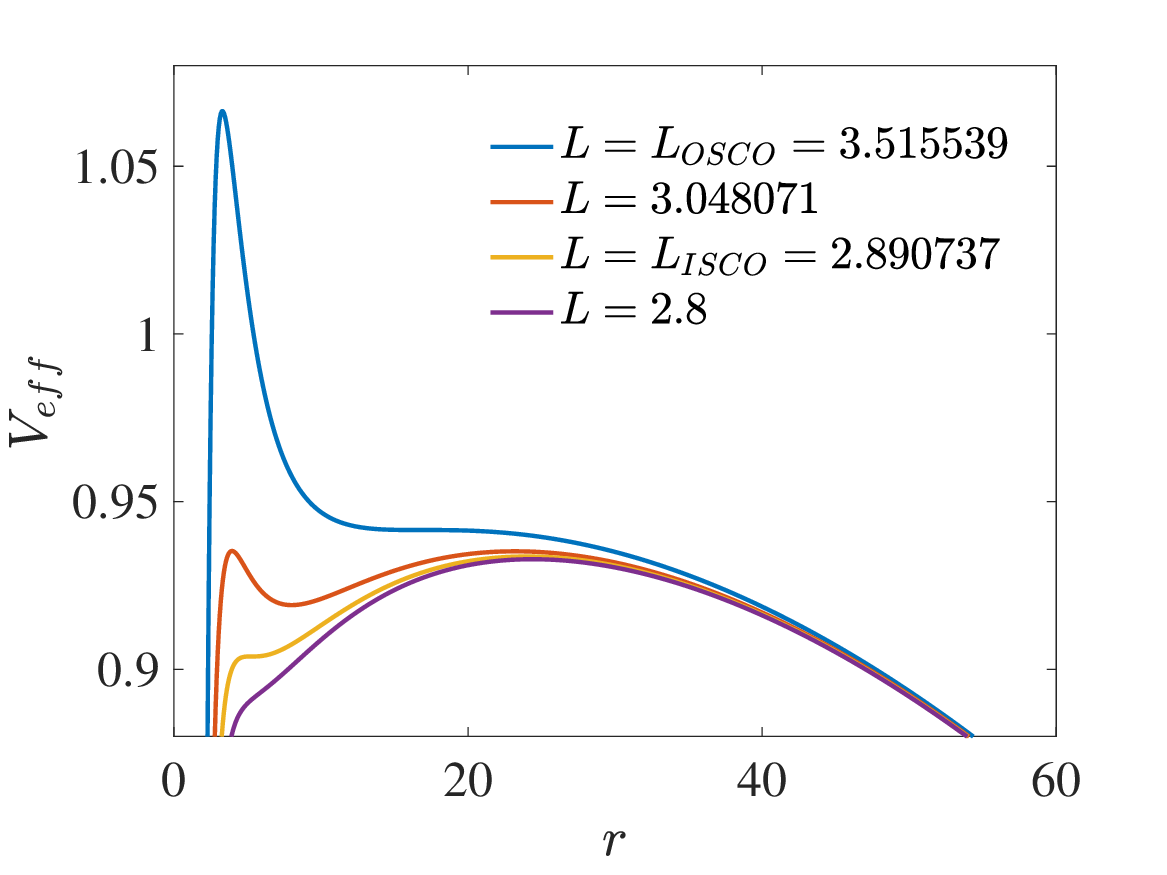}}\label{fig4b}}
\caption{Variation of the effective potential for various values of the angular momentum}
\label{fig4}
\end{figure}

The equation (\ref{10}) serves as the starting point for the subsequent analysis, where the effective potential of timelike particles is expressed as:
\begin{align}
V_{eff}\left(r\right)=\left(1-\frac{2M}{r}-\frac{\Lambda}{3}r^2+\frac{\lambda}{r}\ln\frac{r}{\lambda}\right)\left(1+\frac{L^2}{r^2}\right).\label{16}
\end{align}
By selecting appropriate parameter values, the curves of effective potential varying with radius are shown in Figure \ref{fig4}. According to this figure and the trend of the curve, the effective potential energy curve has at least one extremum, possibly two, and at most three extrema. These points correspond to the stable or unstable circular orbits of the particle, commonly referred to as marginally bound orbits (MBOs). In particular, when the angular momentum $L$ equals $L_\mathrm{ISCO}=3.353590$, there exists an innermost stable circular orbit (ISCO) in spacetime---that is, the circular orbit closest to the black hole that still maintains long-term stability. When the angular momentum $L$ equals $L_\mathrm{OSCO}=4.072766$, the outermost stable circular orbit (OSCO) exists, representing the maximum radius at which a stable circular orbit can occur. For stars orbiting supermassive black holes, since the mass of the star is much less than that of the supermassive black hole, the stars' motion can be approximated as that of test particles moving within the curved spacetime background. Specifically, their trajectories correspond to timelike geodesics, which manifest as bound orbits around the black hole. Accordingly, a detailed analysis of the bound orbits between ISCOs and OSCOs is presented in the following discussion. Furthermore, regardless of the parameter values, the greater the particle's orbital angular momentum, the higher the effective potential energy.

\begin{figure}[H]
\centering
\subfigure[$L=3.353590$]{
\centering{\includegraphics[scale=0.35]{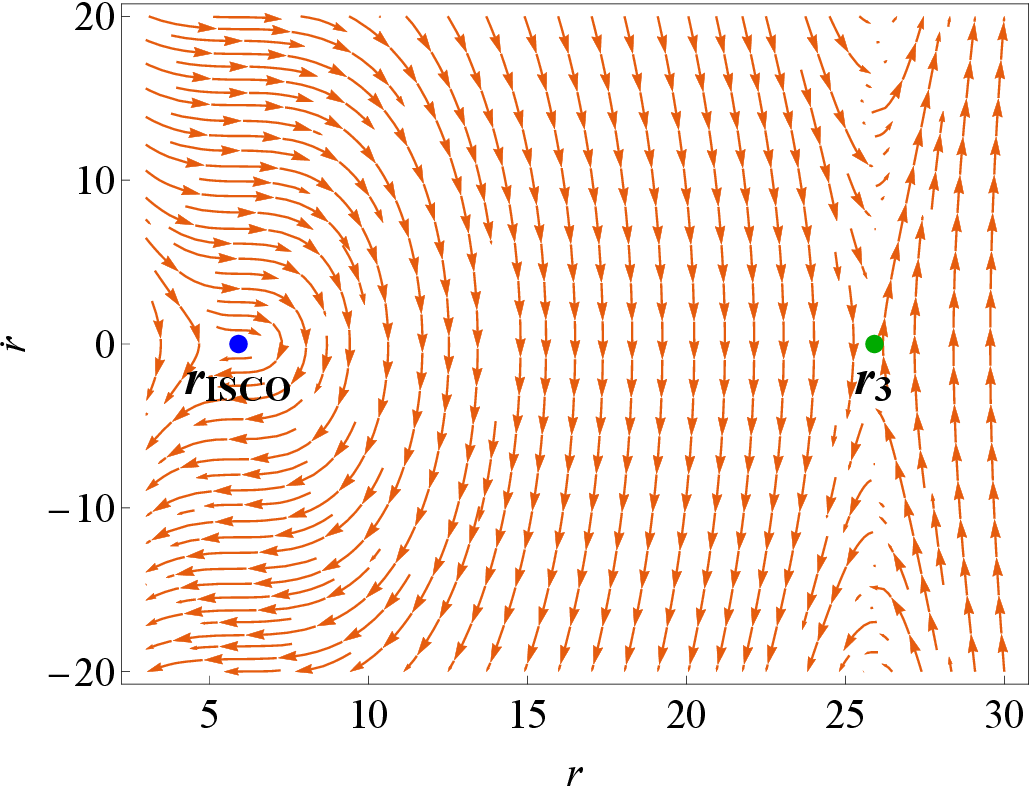}}\label{}}
\subfigure[$L=3.5$]{
\centering{\includegraphics[scale=0.35]{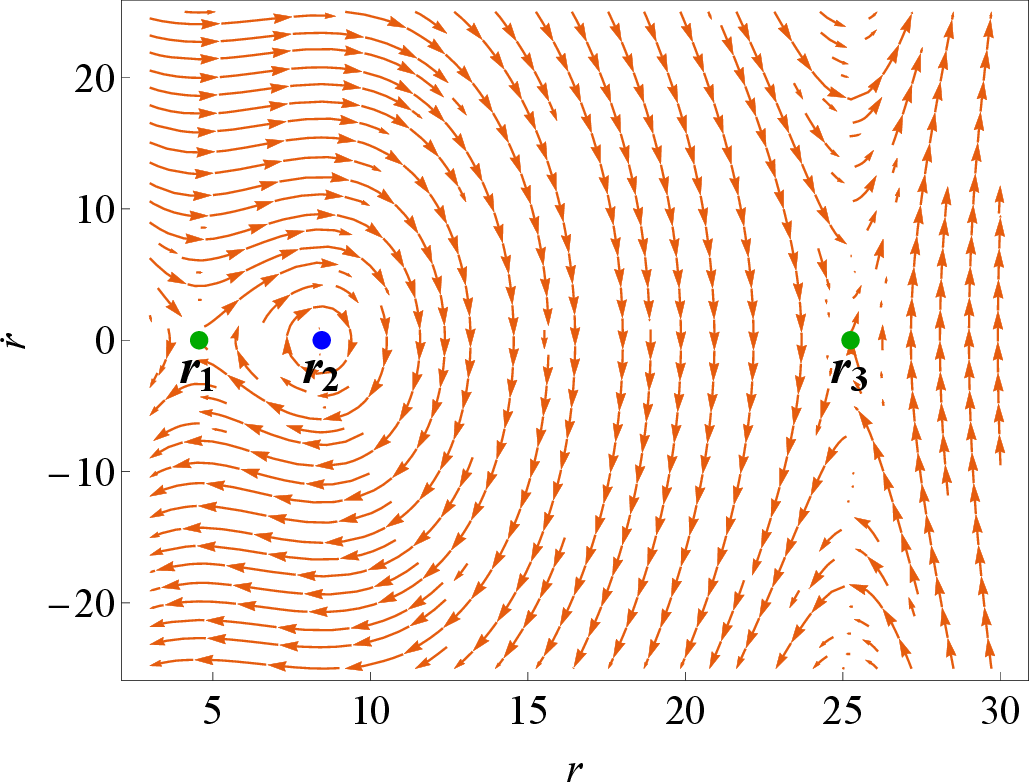}}\label{}}
\subfigure[$L=3.8$]{
\centering{\includegraphics[scale=0.35]{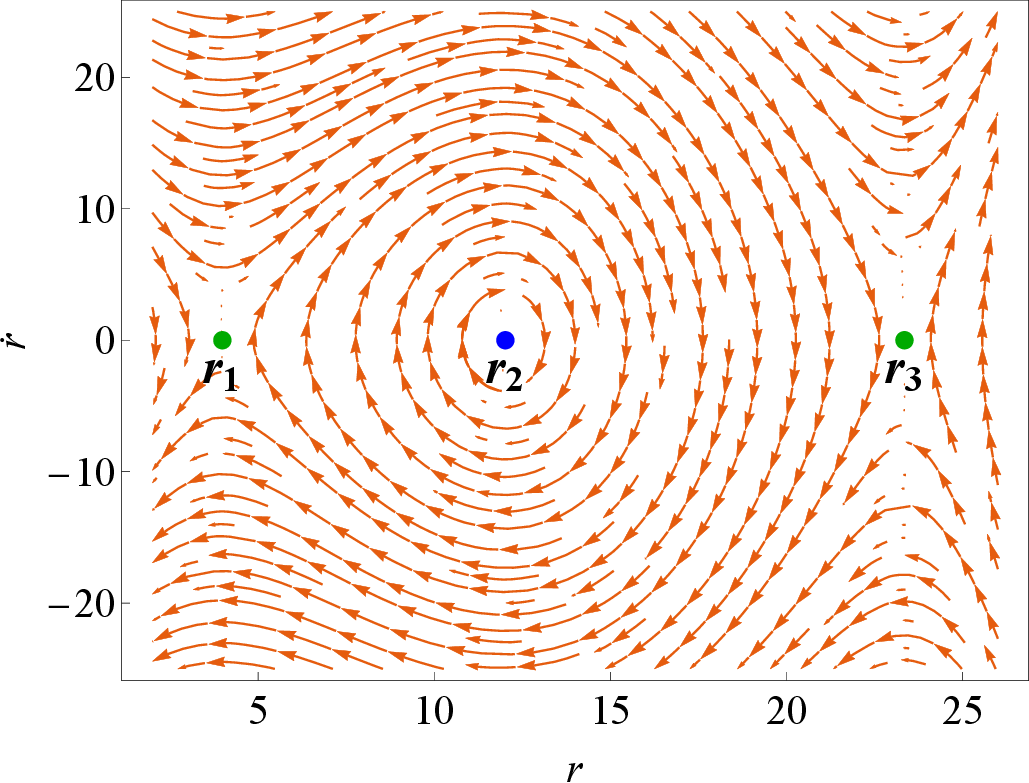}}\label{}}
\subfigure[$L=4.072766$]{
\centering{\includegraphics[scale=0.35]{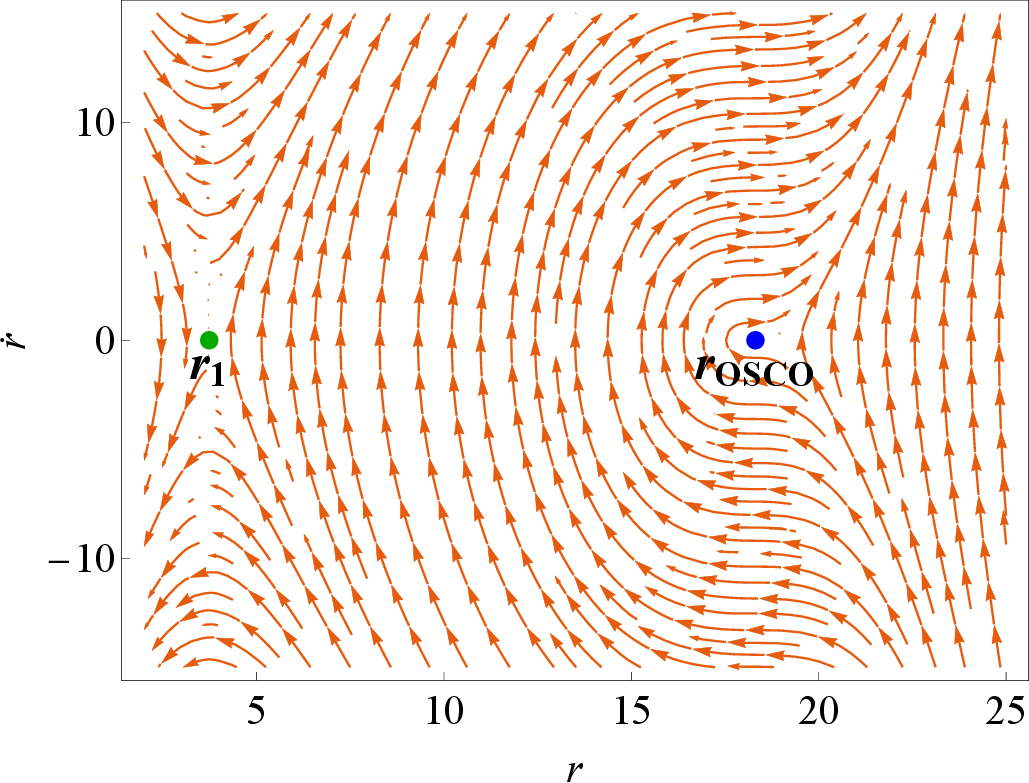}}\label{}}
\caption{Phase portrait of system (\ref{14}): Stable and unstable circular orbits as fixed points.}
\label{fig10}
\end{figure}

To facilitate an intuitive understanding of the long-term dynamical behavior of system (\ref{14}) under given initial conditions, the phase portrait of the system is constructed based on equation (\ref{16}) and the preceding analysis, as presented in Figure \ref{fig10}. In this figure, the points $r_i$ (with $i = 1, 2, 3, \mathrm{ISCO}, \mathrm{OSCO}$) are fixed points satisfying $\dot{r}\left(r_{i}\right) = 0$, and their values correspond to the equilibrium states of the system. Specifically, $r_1$ and $r_3$ are saddle points, characterizing the radii of unstable circular orbits, while $r_2$, $r_\mathrm{ISCO}$, and $r_\mathrm{OSCO}$ are center-type fixed points, characterizing the radii of stable circular orbits. Here, $r_\mathrm{ISCO}$ and $r_\mathrm{OSCO}$ denote the radius of the innermost stable circular orbit and that of the outermost one, respectively.

\begin{figure}[H]
\centering
\subfigure[$\Lambda=0.0001$]{
\centering{\includegraphics[scale=0.35]{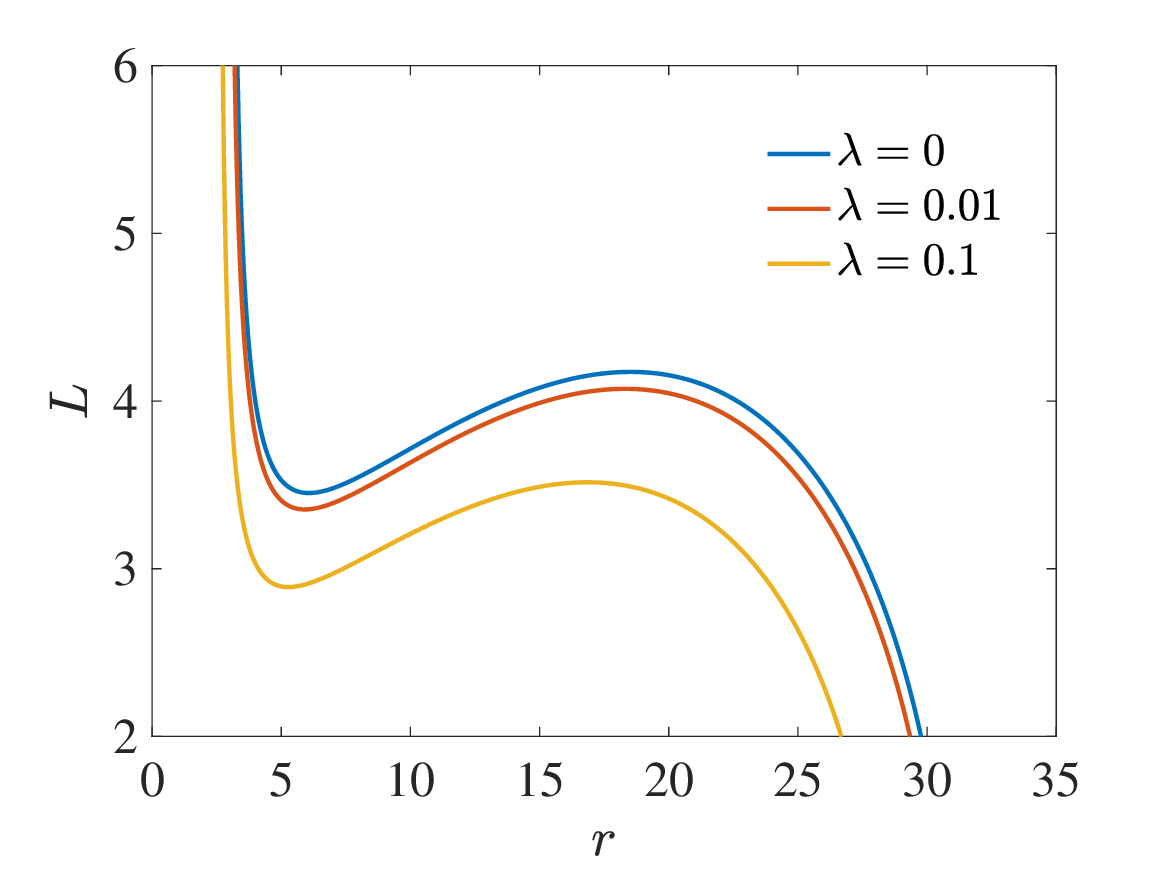}}\label{fig5a}}
\subfigure[$\lambda=0.01$]{
\centering{\includegraphics[scale=0.35]{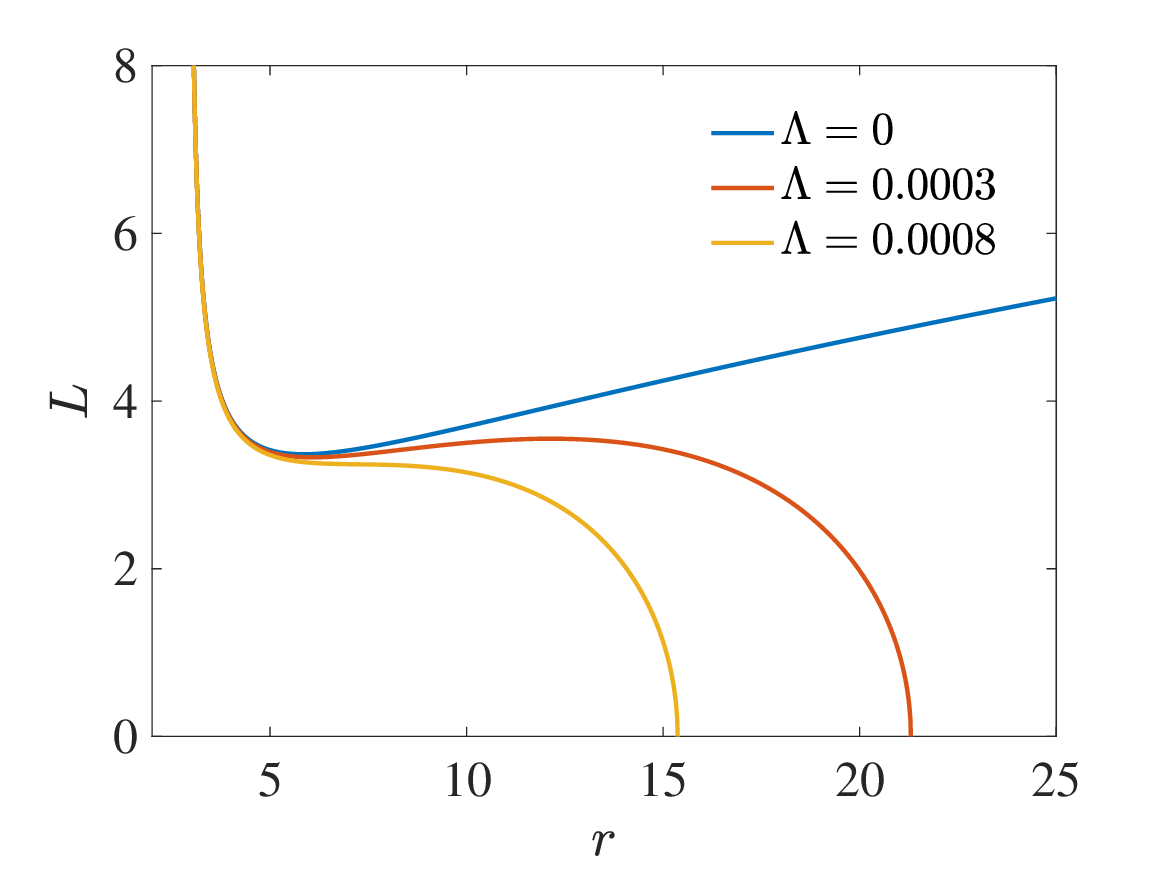}}\label{fig5b}}
\subfigure[$\Lambda=0.0001$]{
\centering{\includegraphics[scale=0.35]{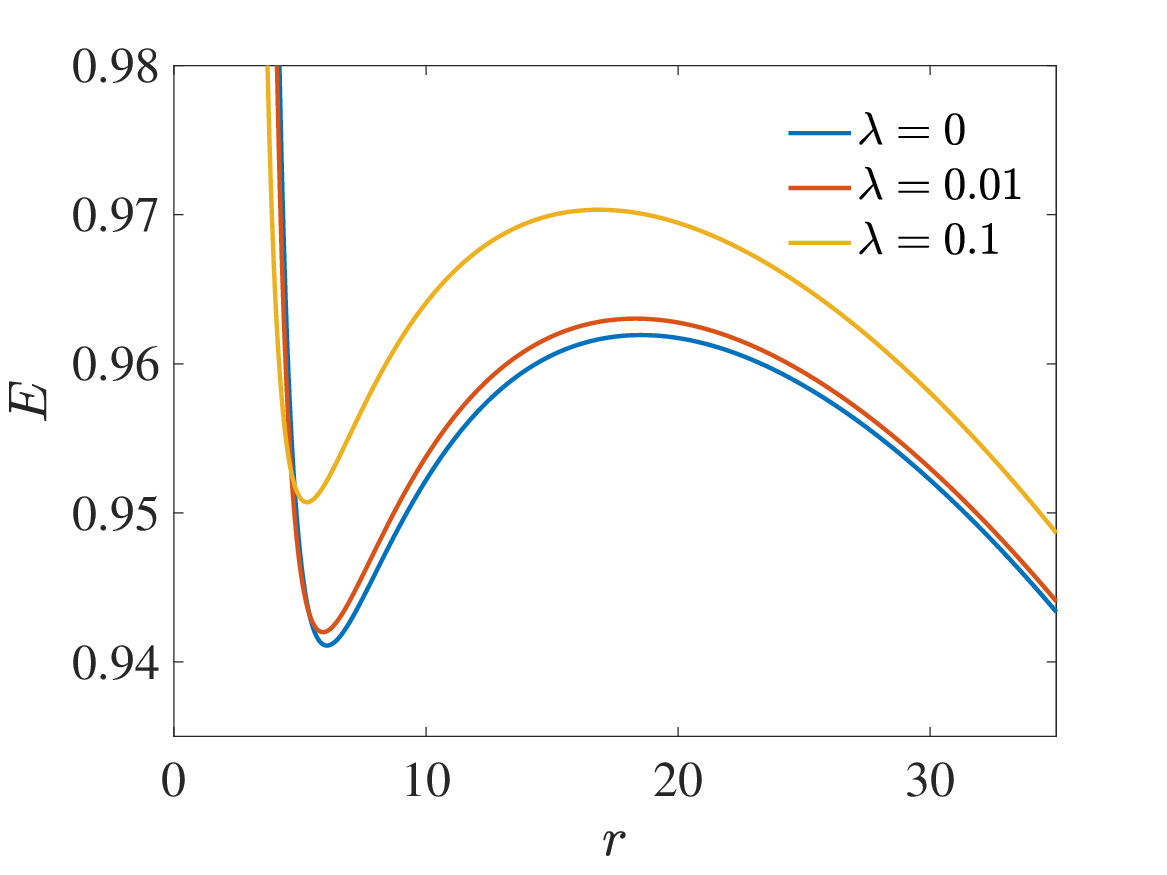}}\label{fig5c}}
\subfigure[$\lambda=0.01$]{
\centering{\includegraphics[scale=0.35]{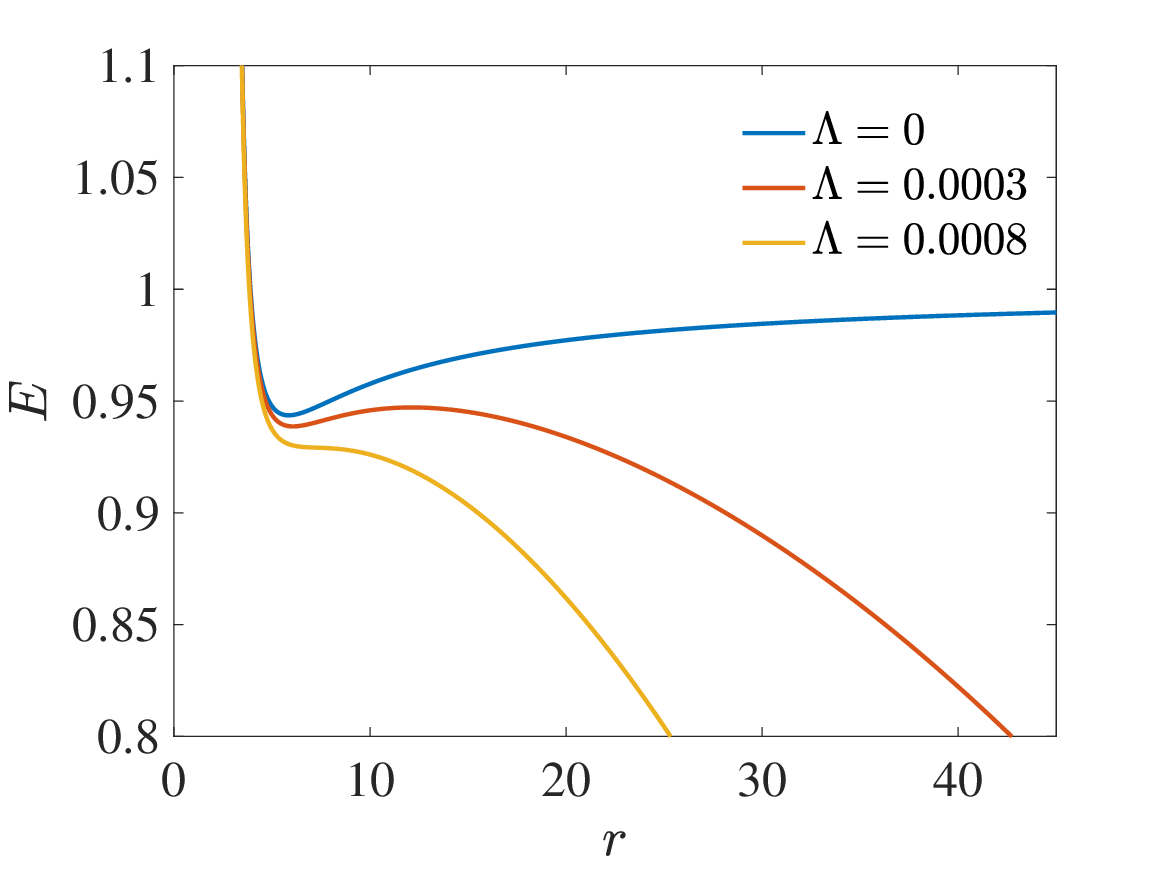}}\label{fig5d}}
\caption{Variation of the angular momentum and the energy for various values of parameters $\lambda$ and $\Lambda$.}
\label{fig5}
\end{figure}

For the MBOs of the timelike particles, the condition satisfies
\begin{align}
V_{eff}\left(r\right)=E^{2}, \qquad \frac{{\rm d}V_{eff}\left(r\right)}{{\rm d}r}=0.\label{17}
\end{align}
Substituting equation (\ref{16}) into the above equation yields
\begin{align}
L^{2}&= \frac{r^2}{3} \left(-3 \lambda -6 M+2 \Lambda  r^3+3 \lambda \ln \frac{r}{\lambda}\right) \left(\lambda +6 M-2 r-3 \lambda \ln \frac{r}{\lambda}\right)^{-1},\label{18}\\
E^{2}&= \frac{r^2}{9} \left(-\frac{2 M}{r}-\frac{\Lambda  r^2}{3}+\frac{\lambda}{r}\ln \frac{r}{\lambda}+1\right)  \nonumber\\
&\quad \times \left( \frac{9}{r^2} + \left(3 \lambda +6 M-2 \Lambda  r^3-3 \lambda \ln \frac{r}{\lambda}\right)^2 \left(\lambda +6 M-3 \lambda  \ln \frac{r}{\lambda}-2 r\right)^{-2} \right).\label{19}
\end{align}
The curves of particle orbital angular momentum and energy varying with radial radius, obtained through numerical simulation, are shown in Figure \ref{fig5} under different parameter values of $\lambda$ and $\Lambda$.

\begin{figure}[H]
\centering
\subfigure[$\Lambda=0.0001$]{
\centering{\includegraphics[scale=0.35]{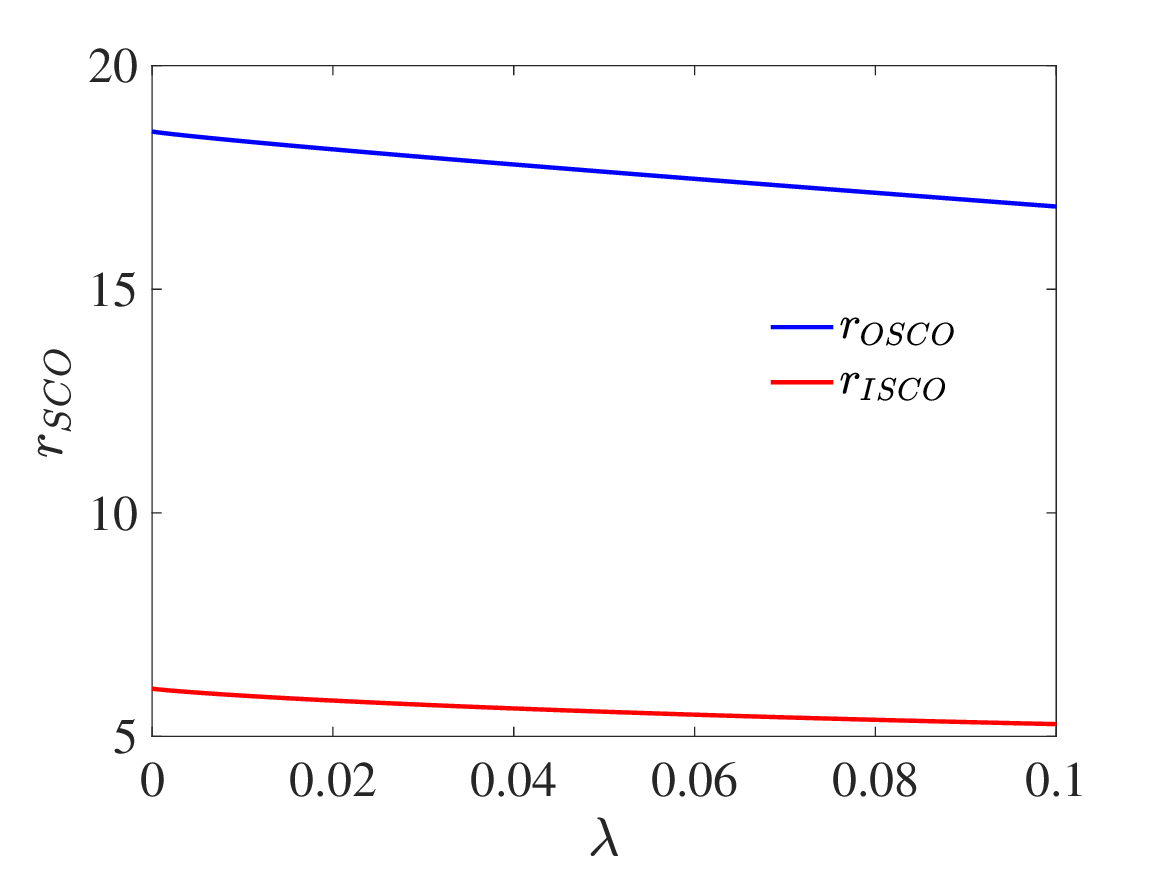}}\label{fig6a}}
\subfigure[$\lambda=0.01$]{
\centering{\includegraphics[scale=0.35]{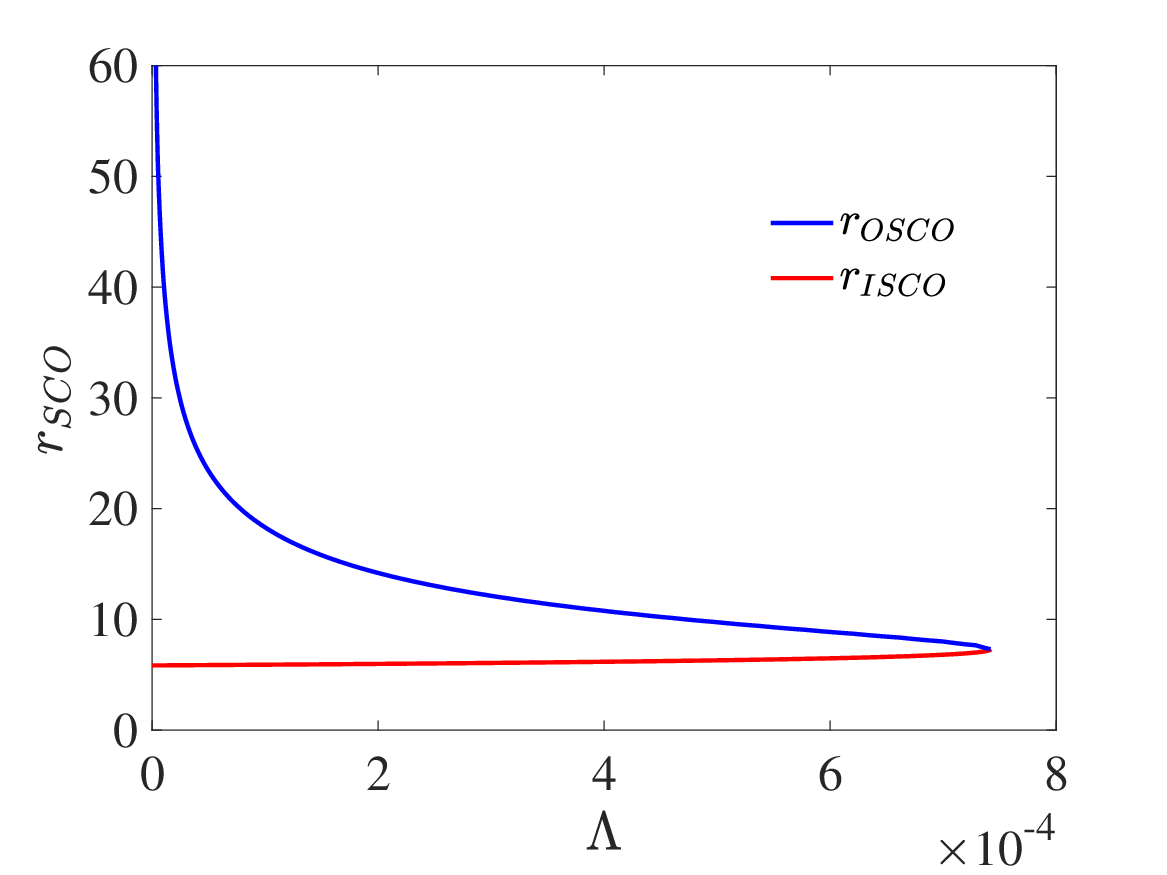}}\label{fig6b}}
\caption{Variation of the radii of ISCO and OSCO for various values of parameters $\lambda$ and $\Lambda$.}
\label{fig6}
\end{figure}

\begin{figure}[H]
\centering{\includegraphics[scale=0.35]{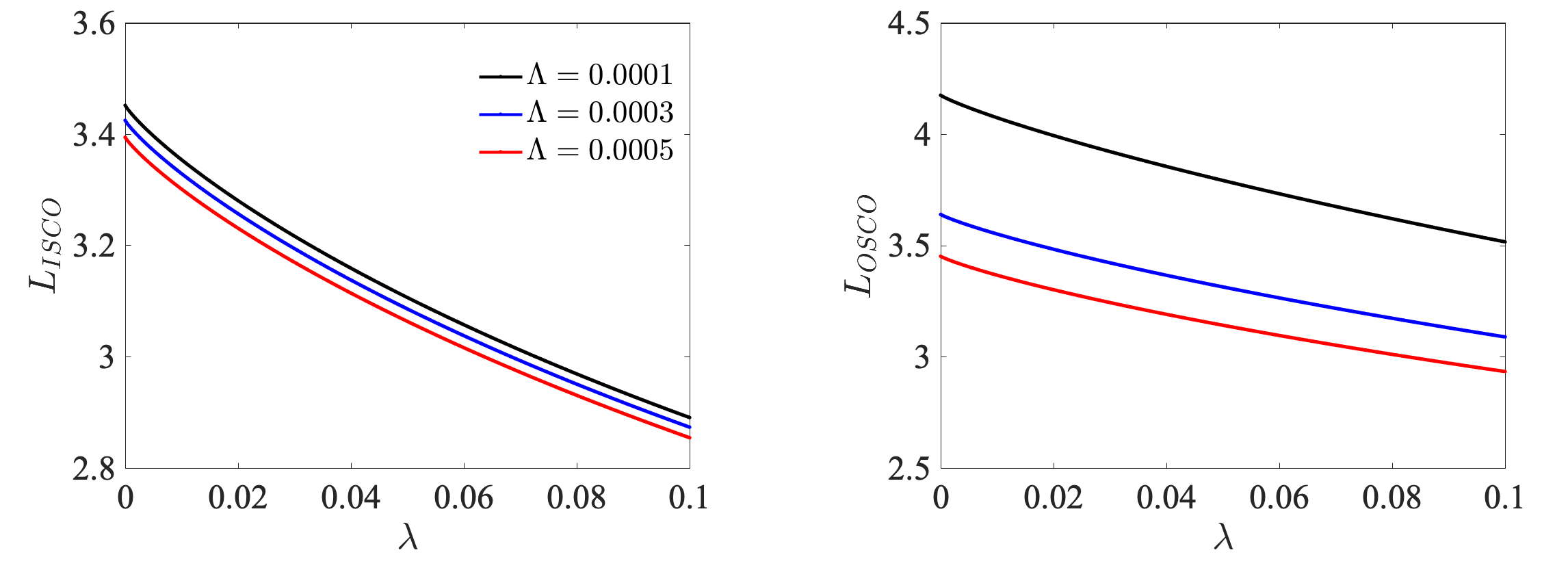}}
\caption{Variation of the angular momentum of ISCO and OSCO for various values of parameters $\lambda$ and $\Lambda$.}
\label{fig7}
\end{figure}

\begin{figure}[H]
\centering{\includegraphics[scale=0.35]{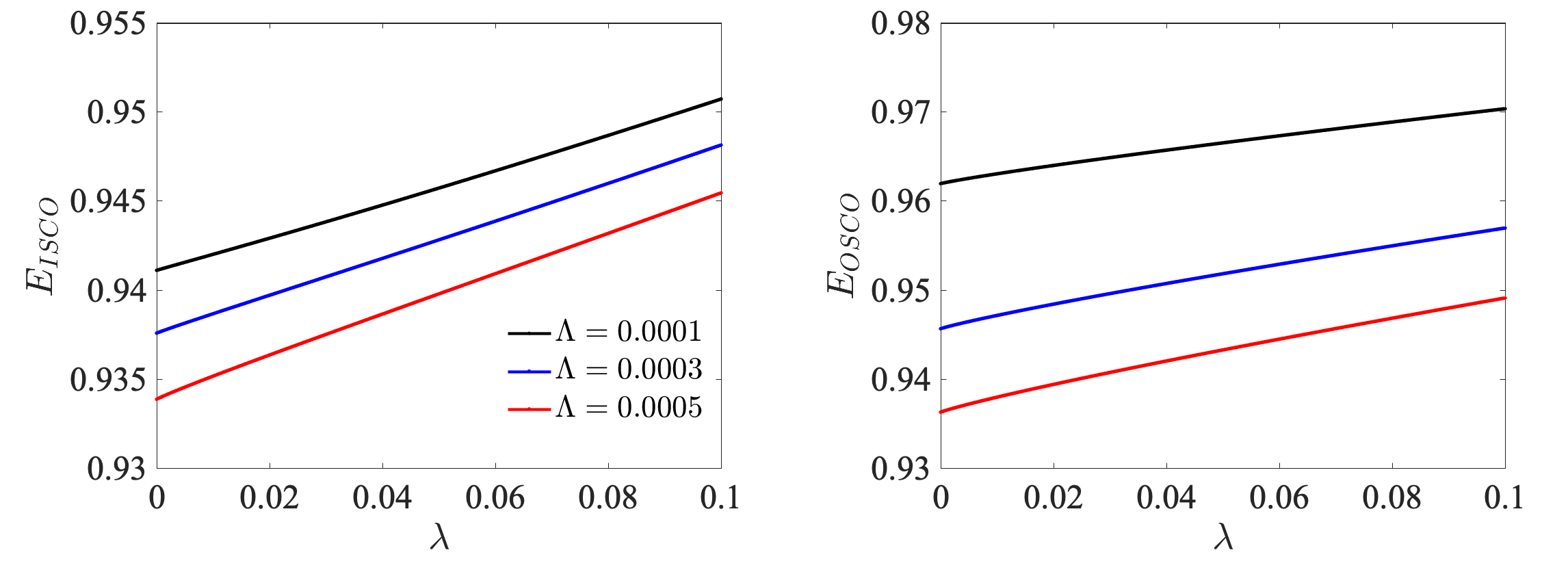}}
\caption{Variation of the energy of ISCOs and OSCOs for various values of the parameters $\lambda$ and $\Lambda$.}
\label{fig8}
\end{figure}

\begin{figure}[H]
\centering{\includegraphics[scale=0.38]{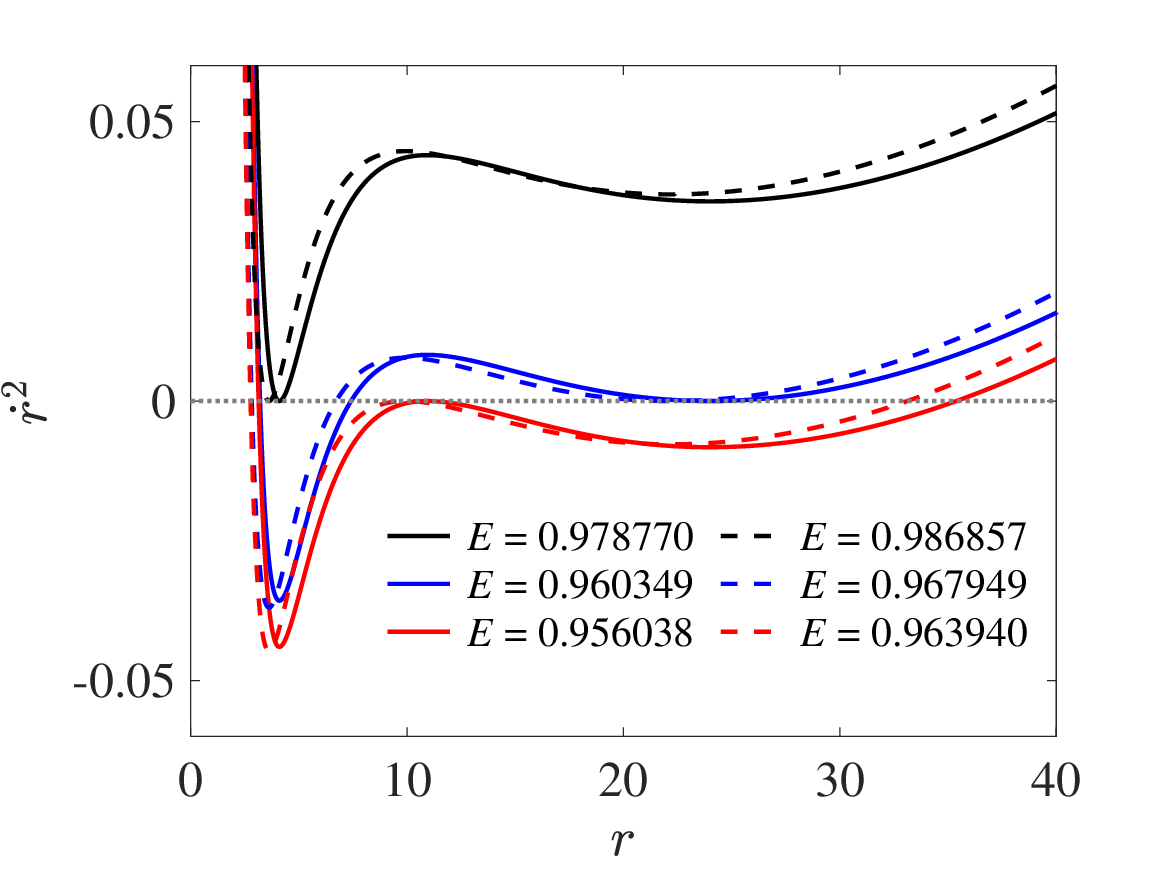}}
\caption{Relationship between ${\dot r}^{2}$ and $r$ for different energies of orbits.}
\label{fig9}
\end{figure}

The condition for the ISCOs/OSCOs satisfies
\begin{align}
V_{eff}\left(r\right)=E^{2}, \qquad \frac{{\rm d}V_{eff}\left(r\right)}{{\rm d}r}=0, \qquad \frac{{\rm d^{2}}V_{eff}\left(r\right)}{{\rm d}r^{2}}=0.\label{20}
\end{align}
i.e.,
\begin{align}
r_{SCO}&=r \left(6 M+\Lambda  r^3-3 r-3 \lambda  \ln \frac{r}{\lambda}\right) \left(-3 \lambda -6 M+2 \Lambda  r^3+3 \lambda  \ln \frac{r}{\lambda}\right) \nonumber\\
& \quad \times \left(6 \lambda ^2+6 M \left(\lambda -4 \Lambda  r^3+2 r\right)+2 \Lambda ^2 r^6+2 \Lambda  r^4-11 \lambda  \Lambda  r^3+9 \lambda  r-3 \lambda \left(\lambda -4 \Lambda  r^3+2 r\right) \ln \frac{r}{\lambda}\right)^{-1},\label{21}\\
L_{SCO}&=\frac{r}{\sqrt{3}}\left(3 \lambda +6 M-2 \Lambda  r^3-3 \lambda  \ln \frac{r}{\lambda}\right)^{1/2}\left(-\lambda -6 M+2 r+3 \lambda \ln \frac{r}{\lambda}\right)^{-1/2},\label{22}\\
E_{SCO}&=\sqrt{\frac{2}{9r}} \left(-6 M-\Lambda  r^3+3 r+3 \lambda  \ln \frac{r}{\lambda}\right)\left(-\lambda -6 M+2 r+3 \lambda  \ln \frac{r}{\lambda}\right)^{-1/2}.\label{23}
\end{align}
Based on equations (\ref{21})-(\ref{23}), the corresponding curves are shown in Figures \ref{fig7}-\ref{fig9}. These figures illustrate the variation of the radius, angular momentum, and energy of ISCO and OSCO with respect to the parameters $\lambda$ and $\Lambda$. As shown in Figure \ref{fig6}, the radius of ISCO decreases as parameter $\lambda$ increases and increases as parameter $\Lambda$ increases. In contrast, the radius of OSCO decreases as both parameters $\lambda$ and $\Lambda$ increase. In particular, when $\Lambda=0.000742$, the ISCO and OSCO merge into one. Similarly, Figure \ref{fig7} shows that increases in parameters $\lambda$ and $\Lambda$ lead to decreases in the angular momentum of ISCO and OSCO. Figure \ref{fig8} illustrates that a larger parameter $\lambda$ corresponds to higher orbital energy, whereas a larger parameter $\Lambda$ corresponds to lower orbital energy.

The discussion of bound orbits is based on equation (\ref{14}), which describes the motion of particles. It is noted that the bound orbits, discussed in this paper, is confined between ISCOs and OSCOs. Therefore, the chosen angular momentum should be $L=L_{ISCO}+\epsilon \left(L_{OSCO}-L_{ISCO}\right)$, where $\epsilon \in [0,\;1]$. Setting $\epsilon=0.5$ yields $L=\left(L_{ISCO}+L_{OSCO}\right)/2$. Figure \ref{fig9} shows the relationship between ${\dot r}^{2}$ and $r$ for different energies of orbits. Among them, the solid line corresponds to the parameter $\lambda$ with a value of $0.01$, and the dashed line corresponds to the parameter with a value of $0.1$. The curve $r-{\dot r}^{2}$ consistently exhibits three extrema: one maximum and two minima. As the orbital energy varies, the function values at these extrema also change. However, not all energy values are considered in this analysis. To study bound orbits between the edge orbits and the inner stable circular orbits, the focus is on cases where the function value at the maximum point is greater than or equal to zero, and the function values at all minimum points are less than or equal to zero. For example, when $\lambda=0.01$ (solid line), the orbital energy ranges from 0.956038 to 0.960349; when $\lambda=0.1$ (dashed line), the orbital energy ranges from 0.963940 to 0.967949. That is, the situation between the red curve and the blue curve.

\begin{figure}[H]
\centering{\includegraphics[scale=0.38]{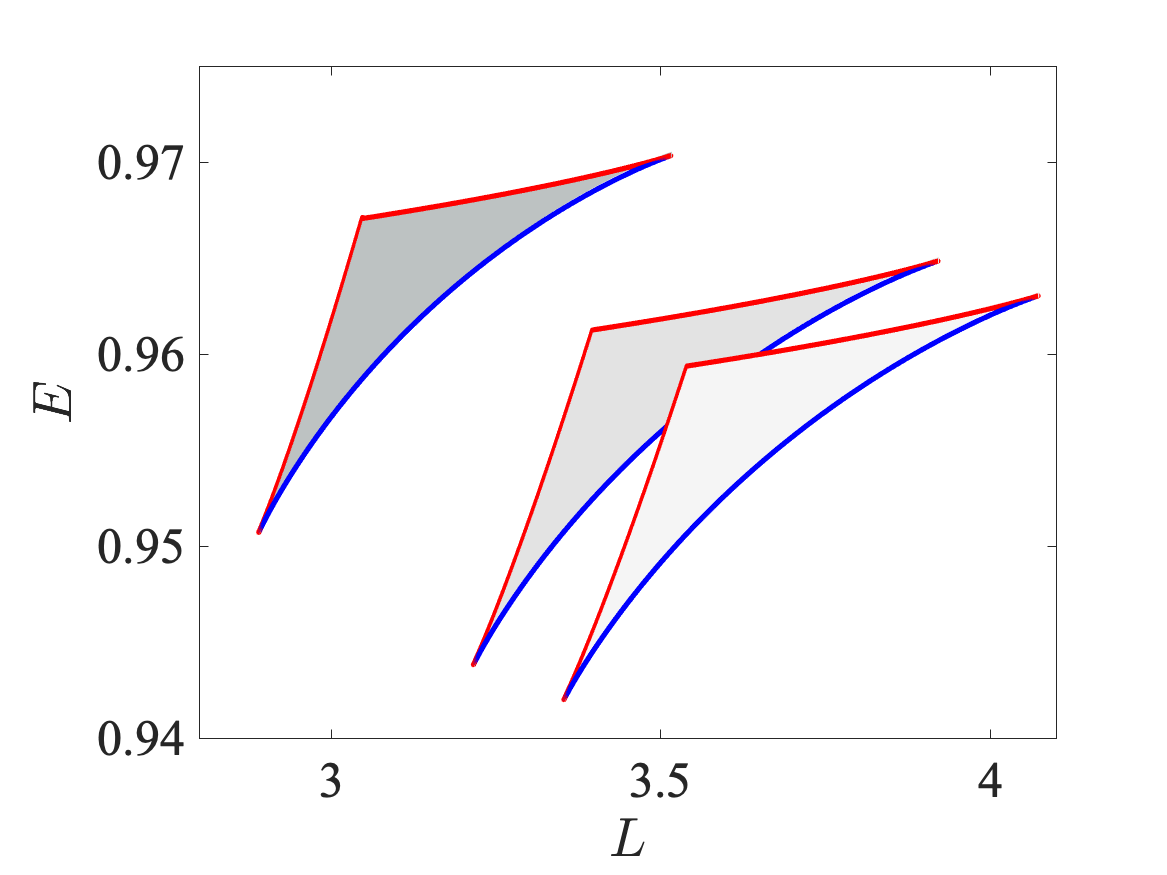}}
\caption{Allowed regions $\left(E,\; L\right)$ of particle motion. The three regions are distinguished by colors that gradually deepen from light to dark, representing three distinct values of the parameter $\lambda$. Specifically, the light, medium, and dark areas correspond to $\lambda=0.01$, $0.03$, and $0.1$, respectively, with darker shades denoting larger parameter values.}
\label{fig11}
\end{figure}

Based on the above analysis, it can be inferred that different angular momenta of the given orbit correspond to distinct ranges of orbital energy. Therefore, the allowable ranges of energy and angular momentum for the particle's orbital motion can be determined, as illustrated in Figure \ref{fig11}. The various deep gray triangular regions in the figure correspond to different values of the parameter $\lambda$. The larger the value of $\lambda$, the higher the position of the region. Additionally, the blue curve in the image represents the energy associated with the particle's inner circular orbit, while the red curve represents the energy corresponding to the MBO.

\section{Orbital pericenter precession, constraints on PFDM parameters from S2 star's motion, periodic orbits and their gravitational wave}\label{se-4}
\subsection{Orbital pericenter precession and constraints on PFDM parameters from S2 star's motion}
In the spacetime of the Schwarzschild-dS BH immersed in PFDM, periodic orbits as a special class of trajectories are of significant interest. By using the orbital classification method \cite{Levin2008, Levin2009a, Levin2009b}, the study of periodic orbits of timelike particles moving between the ISCOs and OSCOs is focused on. Specifically, a rational number $q$, which describes the orbital characteristics of periodic orbits and precessing orbits, is defined through a triple of integers $(z,\;w,\;v)$ as
\begin{align}
q=w+\frac{v}{z}=\frac{\Delta \phi}{2\pi}-1,\label{24}
\end{align}
here $0 \leq v<z$, $v/z$ is an irreducible fraction, and the apsidal angle $\Delta \phi$ denotes the total angular accumulation from one apoapsis to the next over a complete radial period, given by the expression
\begin{align}
\Delta \phi&=\oint {\rm d} \phi \nonumber\\
&=2\int_{r_{pe}}^{r_{ap}} \frac{\dot \phi}{\dot r} {\rm d} r \nonumber\\
&=2\int_{r_{pe}}^{r_{ap}} \frac{L}{r^2 \sqrt{E^2-\left(1-\frac{2M}{r}-\frac{\Lambda}{3}r^2+\frac{\lambda}{r}\ln\frac{r}{\lambda}\right)\left(1+\frac{L^2}{r^2}\right)}} {\rm d} r,\label{25}
\end{align}
where, $r_{pe}$ and $r_{ap}$ represent the orbital turning points (the periapsis and apoapsis, respectively), lying between the ISCO and OSCO. They are determined by the condition $\dot{r}^2 = E^2-V_{\text{eff}} = 0$. The triple of integers $(z, w, v)$ offers a meaningful geometric interpretation of the orbit’s shape: $z$ corresponds to the number of ``zoom leaves'', $w$ indicates the ``whirl count'', and $v$ (with $0 \leq v < z$) denotes the ``vertex number'', which connects successive apoapsis points along the orbit. Regarding periastron precession, the particle’s trajectory is governed by \cite{Straumann2004}
\begin{align}
r = \frac{a\left(1-e^{2}\right)}{1+ e \cos \Psi},\label{26}
\end{align}
where the turning points correspond to the periapsis $ r_{pe} = a \left(1-e\right) $ and the apoapsis $ r_{ap} = a \left(1+e\right) $. Here, $a$ and $e$ represent the semi-major axis and eccentricity of the orbit, respectively, while $\Psi$ denotes the intersection angle between the semi-major axis and the orbital radial direction. Accordingly, the angle $\phi$ in this case is given by
\begin{align}
\Delta \phi=2\int_{r_{pe}}^{r_{ap}} \frac{{\rm d} \phi}{{\rm d} r} {\rm d} r=2\int_{0}^{\pi} \frac{{\rm d} \phi}{{\rm d} \Psi}{\rm d} \Psi,\label{27}
\end{align}
with
\begin{align}
\frac{{\rm d} \phi}{{\rm d} \Psi}&=\frac{{\rm d} \phi / {\rm d} r}{{\rm d} \Psi / {\rm d} r}\nonumber\\
&=\frac{ae\left(1-e^{2}\right)L\sin\Psi}{r^{2}\left(1+e\cos\Psi\right)^{2}\sqrt{E^2-\left(1-\frac{2M}{r}-\frac{\Lambda}{3}r^2+\frac{\lambda}{r}\ln\frac{r}{\lambda}\right)\left(1+\frac{L^2}{r^2}\right)}}.\label{28}
\end{align}
Given the conditions $\dot{r}|_{r=r_{pe}} = 0$ and $\dot{r}|_{r=r_{ap}} = 0$, the orbital angular momentum $L$ and energy $E$ can be expressed as
\begin{align}
E^{2} & =\frac{\left(r_{pe}^{2}-r_{ap}^{2}\right) f\left(r_{pe}\right) f\left(r_{ap}\right)}{r_{pe}^{2} f\left(r_{ap}\right)-r_{ap}^{2} f\left(r_{pe}\right)}, \label{29}\\
L^{2} & =\frac{r_{pe}^{2} r_{ap}^{2}\left(f\left(r_{pe}\right)-f\left(r_{ap}\right)\right)}{r_{pe}^{2} f\left(r_{ap}\right)-r_{ap}^{2} f\left(r_{pe}\right)}.\label{30}
\end{align}
Then equation (\ref{28}) can be rewritten by introducing the dimensionless parameters $\Lambda^{*} = \Lambda M^2$ and $\lambda^{*}= \lambda / M$ as
\begin{align}
\frac{{\rm d} \phi}{{\rm d} \Psi}=&1+\frac{\left(3+e \cos \psi \right)M}{a\left(1-e^2\right)}+\lambda^{*}\left(\frac{\left(3+e \cos \psi \right)M}{2 a \left(e^2-1\right)}-\frac{3 \left(3+e \cos \psi \right)^2 M^2}{2 a^2 \left(e^2-1\right)^2 }\right)\nonumber\\
&+\Lambda^{*}\left(\frac{a^3 \left(e^2-1\right) \left(e^2-2 e \cos \psi -3\right)}{6 M^{3} \left(1+e \cos \psi\right)^2}+\frac{a^2 \left(e^2-2 e \cos \psi -3\right) \left(e^2-3 e \cos \psi -6\right)}{6 M^{2} \left(1+e \cos \psi\right)^2}\right)+\cdots.\label{31}
\end{align}
Substituting the above equation into equation (\ref{27}) yields the precession angle as
\begin{align}
\Delta \omega_{PFDM}=&\Delta \phi -2\pi \nonumber\\
=&\frac{6 \pi M}{a\left(1-e^2\right)} -\lambda^{*} \left(\frac{3 \pi  M}{a \left(1-e^2\right)}+\frac{3 \pi \left(e^2+18\right) M^2}{2 a^2 \left(e^2-1\right)^2}\right) \nonumber\\
&+\Lambda^{*} \left(\frac{\pi  a^3 \left(1-e^2\right)^{1/2}}{M^3}+\frac{2 \pi  a^2 \left(2-e^2+\left(1-e^2\right)^{1/2}\right)}{\left(1-e^2\right)^{1/2} M^2}\right).\label{32}
\end{align}
In the limit $\Lambda^{*} \to 0$, $\lambda^{*} \to 0$, this result reduces to the general relativistic expression for Schwarzschild black hole, $\Delta \omega_{GR} = \Delta\phi - 2\pi = 6\pi M / \left[a\left(1-e^2\right)\right]$ \cite{Weinberg1972}.

Recently, the GRAVITY collaboration reported a measurement of the GR Schwarzschild Precession (SP) for the S2 star orbiting Sgr A* using combined spectroscopic and astrometric data \cite{Abuter2020}. The ratio of the observed precession rate to the value predicted by GR was found to be
\begin{align}
f_{SP} \equiv \frac{\Delta \omega_{S2}}{\Delta \omega_{GR}}=1.1 \pm 0.19.\label{33}
\end{align}
Here, $\Delta \omega_{S2}$ represents the observed precession of the S2 star, and $\Delta \omega_{GR}$ represents the general relativistic prediction. The clear deviation of this ratio from unity indicates a potential window for alternative gravitational theories beyond GR and can therefore be used to constrain model parameters. Therefore, the central supermassive black hole in Sgr A* is assumed to be a Schwarzschild-dS BH immersed in PFDM, and the precessional motion of the S2 star's orbit is taken to follow the trajectory described above. Accordingly, the precession shift $f_{SP}$ is given by
\begin{align}
f_{SP} \equiv \frac{\Delta \omega_{PFDM}}{\Delta \omega_{GR}}=&1-\lambda^{*} \left(\frac{1}{2}+\frac{\left(e^2+18\right) M}{4 a \left(1-e^2\right)}\right)\nonumber\\
&+\Lambda^{*} \left(\frac{a^4 \left(1-e^2\right)^{3/2}}{6 M^4}+\frac{a^3 \left(1-e^2\right)^{1/2} \left(2 -e^2+\left(1-e^2\right)^{1/2}\right)}{3 M^3}\right).\label{34}
\end{align}

\begin{figure}[H]
\centering{\includegraphics[scale=0.3]{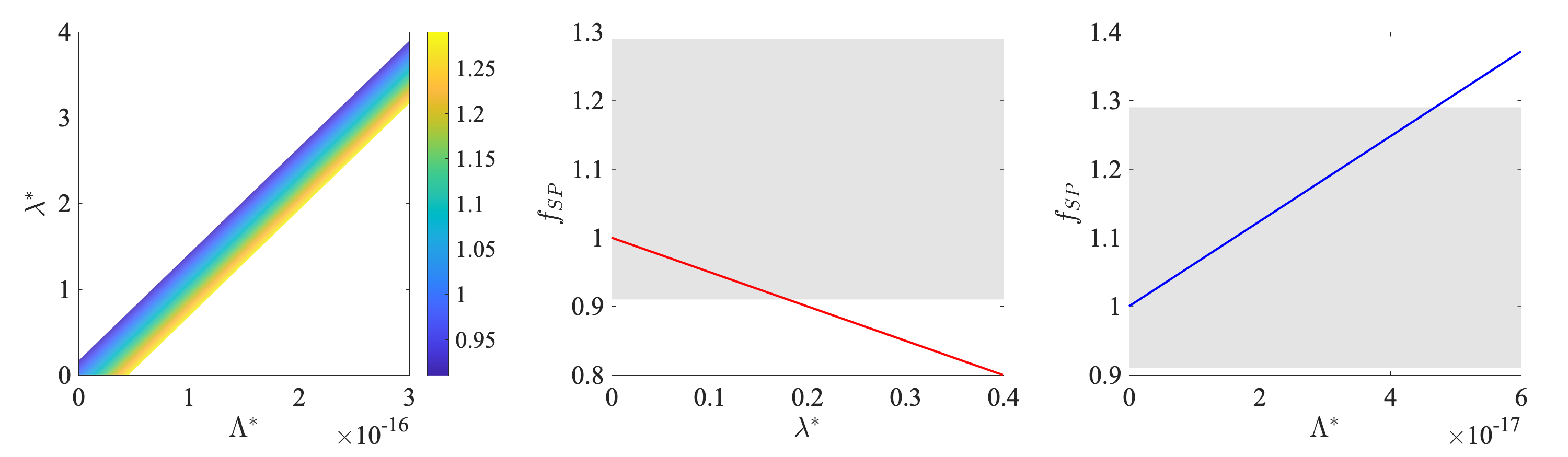}}
\caption{Allowed region of the $\left(\Lambda^{*}, \lambda^{*}\right)$ parameter space under the observational constraint on the ratio (left panel, gradient-colored band), and the $f_{SP}$ ratio as versus the parameters $\Lambda^{*}$ and $\lambda^{*}$ (middle and right panels).}
\label{fig12}
\end{figure}

Using the observed orbital parameters of the S2 star, a black hole mass $M_{\text{Sgr A}} = 4.3 \times 10^6$ M$_{\odot}$, distance $D = 8.35\ \text{kpc}$, semi-major axis \(a = 125.058\ \text{mas}$, and eccentricity $e = 0.884649$, Figure \ref{fig12} presents the allowed region of the $\left(\Lambda^{*}, \lambda^{*}\right)$ parameter space under the observational constraint $1.1_{-0.19}^{+0.19}$ on this ratio (left panel, gradient-colored band), together with the curves of the $f_{SP}$ ratio as a function of the dimensionless parameters $\Lambda^{*}$ and $\lambda^{*}$ (middle and right panels). 

The analysis leads to the constraint $0 \leq \lambda^{*} \leq 0.179685$, which corresponds to the physical bound of $0 < \lambda < 7.726412 \times 10^{5}$ M$_{\odot}$. It indicates that the current measurement of the S2 star’s orbital precession cannot rule out the scenario of a black hole immersed in PFDM and dark energy. Furthermore, these measurements place stringent limits on the possible influence of the dark matter density profile and the dark energy equation of state on the black hole spacetime in realistic astrophysical environments.

\subsection{Periodic orbits of the timelike particles}
\begin{figure}[H]
\centering
\subfigure[$L=\left(L_{ISCO}+L_{OSCO}\right)/2$]{
\centering{\includegraphics[scale=0.38]{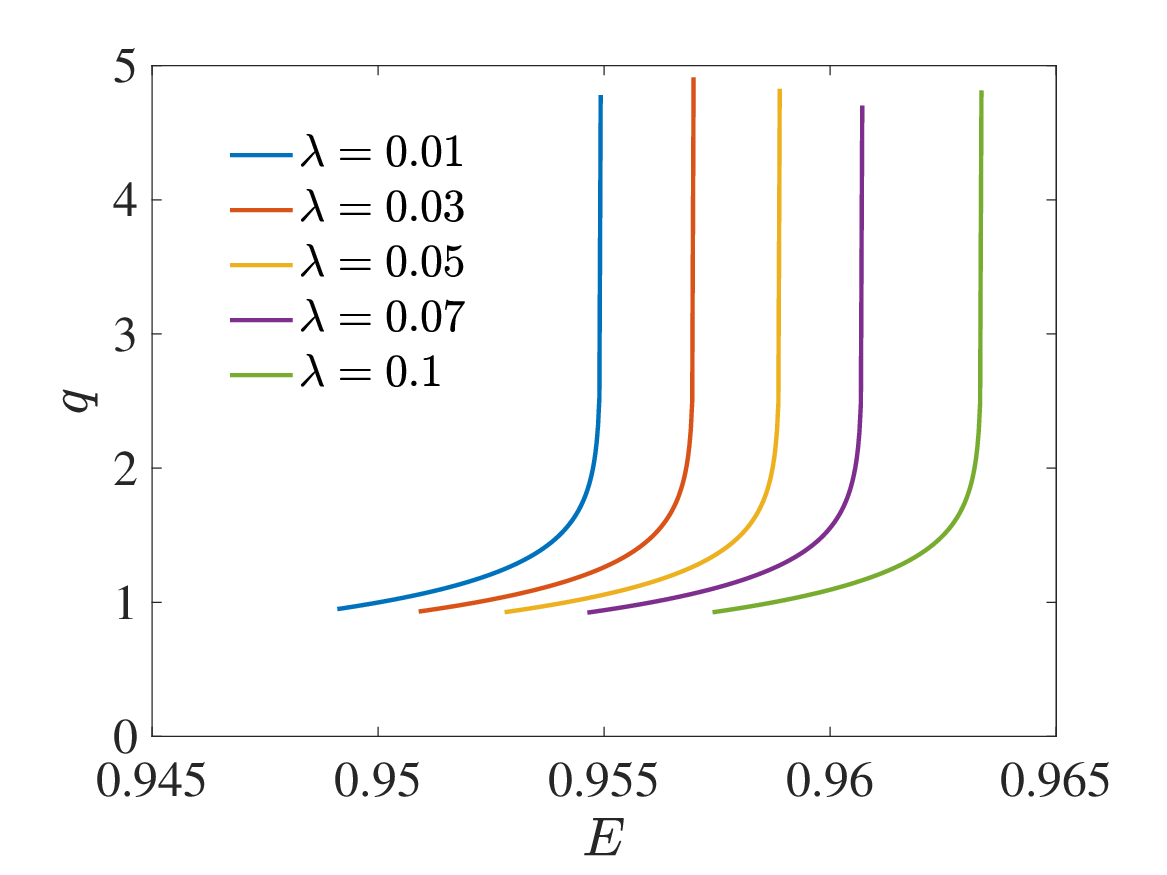}}\label{fig13a}}
\subfigure[$E=0.96$]{
\centering{\includegraphics[scale=0.38]{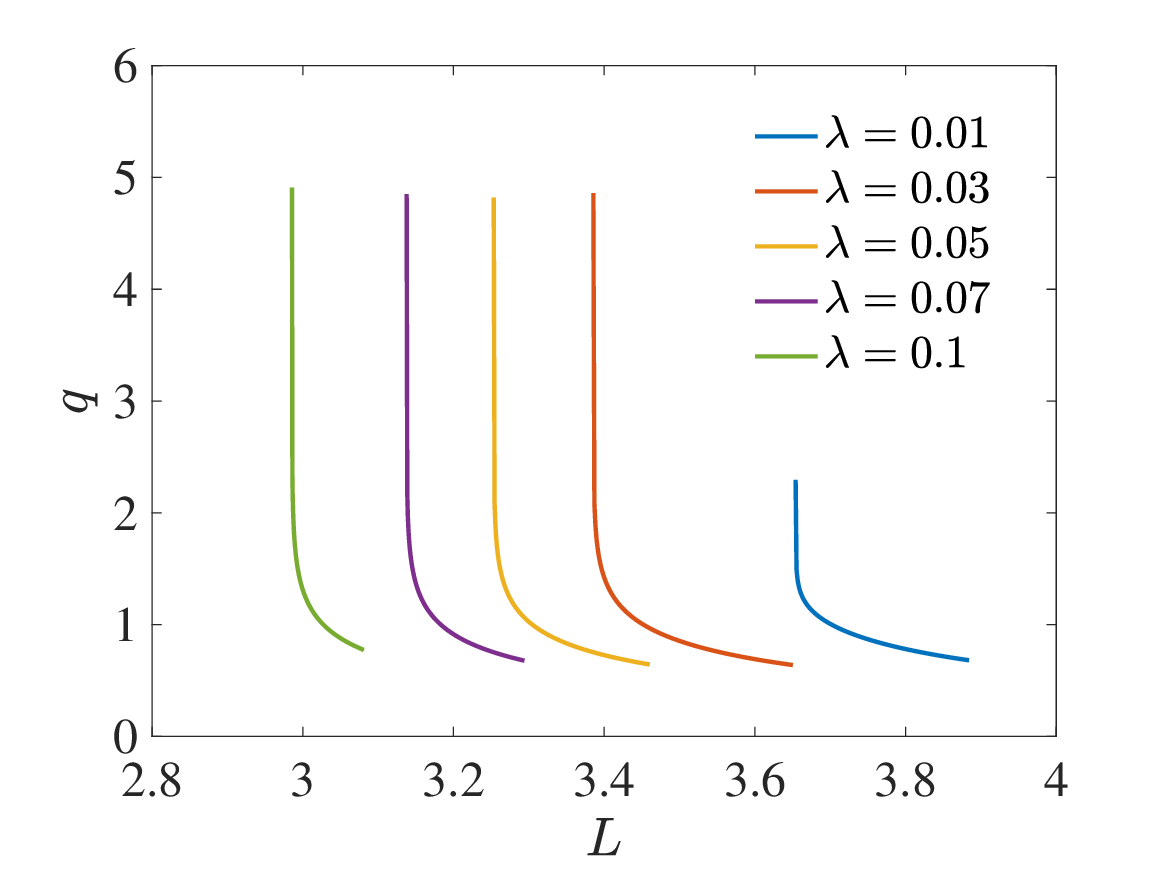}}\label{fig13b}}
\caption{Variation of the zoom-whirl parameter $q$ with the parameter $\lambda$, orbital energy, and angular momentum.}
\label{fig13}
\end{figure}

Focus on equations (\ref{25}) and (\ref{26}), the influence of different parameter values on the zoom-whirl parameter $q$ in a periodic orbit is illustrated in Figure \ref{fig13}, specifically through variations in orbital energy and angular momentum. In Figure \ref{fig13a}, the angular momentum is fixed to $L=\left(L_{ISCO}+L_{OSCO}\right)/2$ , and the effect of varying energy on $q$ is shown for different values of the parameter $\lambda$. The results indicate that as energy increases, $q$ gradually increases and eventually approaches infinity. Moreover, with increasing $\lambda$, the curve shifts to the right. This suggests that, with the zoom-whirl number held constant, maintaining the same dynamical behavior of the periodic orbit requires a higher energy as $\lambda$ increases. Figure \ref{fig13b} fixes the energy at $E=0.96$ and presents the effect of angular momentum variations on $q$ under different $\lambda$ values. The observed trend is opposite to that in Figure \ref{fig13a}. Specifically, for a fixed $\lambda$, a smaller angular momentum leads to a larger value of $q$, which also approaches infinity. Meanwhile, the curve shifts to the left, indicating that, for a given zoom-whirl number, a larger $\lambda$ reduces the angular momentum required to sustain the same dynamical behavior of the periodic orbit. The analysis of the parameter $q$ in zoom-whirl periodic orbits yields results analogous to those obtained from the energy-angular momentum $(E, L)$ region diagram (Figure \ref{fig11}), implying that the two descriptions convey equivalent physical information.

\begin{figure}[H]
\centering{\includegraphics[scale=0.58]{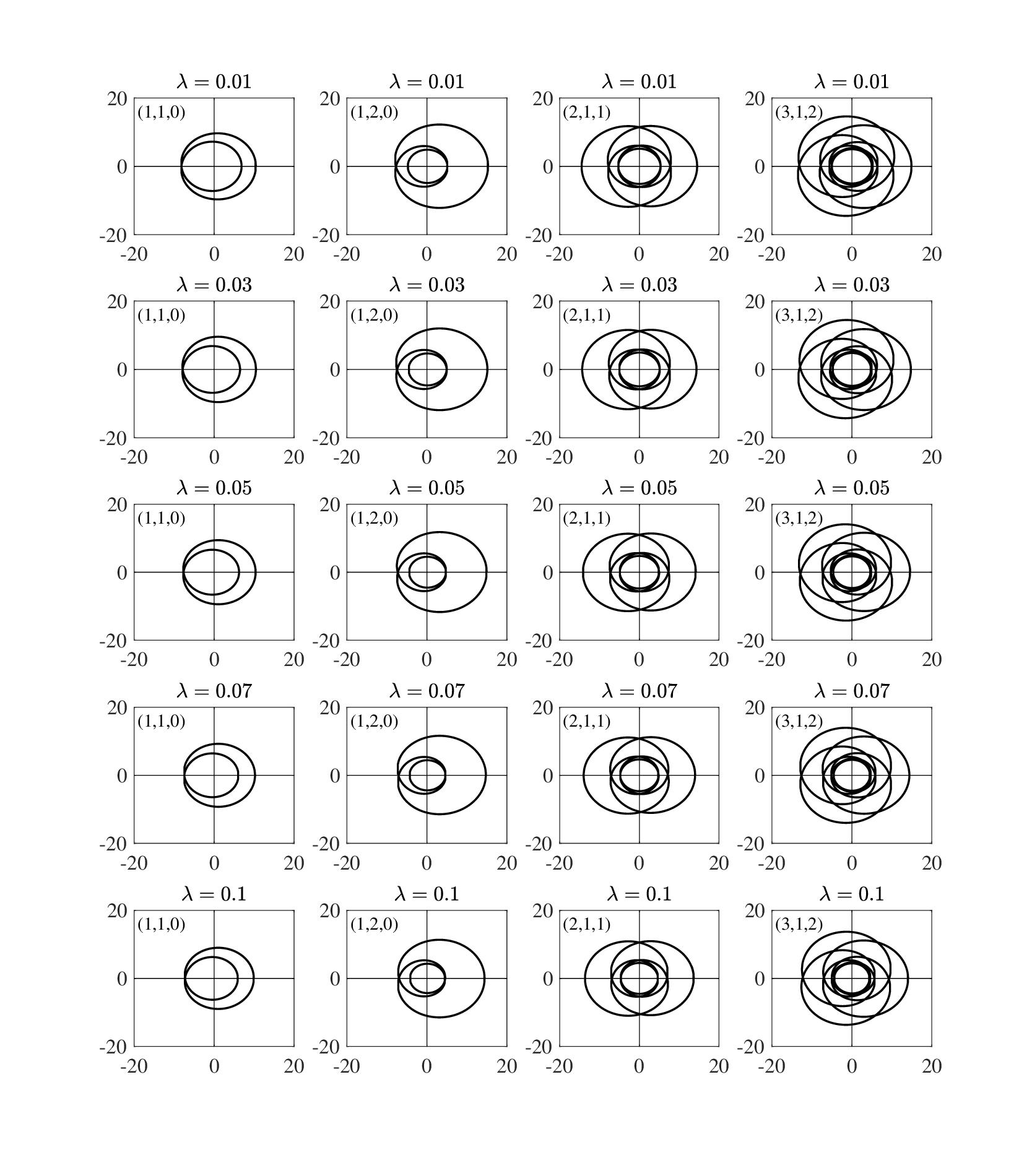}}
\caption{The zoom-whirl periodic orbits of timelike particles between the ISCO and OSCO.}
\label{fig14}
\end{figure}

Figure \ref{fig14} shows the zoom-whirl periodic orbits of the test particle between the ISCO and OSCO around the Schwarzschild-dS BH immersed in PFDM. It consists of twenty subgraphs arranged in five rows and four columns, systematically illustrating the influence of different parameter combinations on the behavior of zoom-whirl periodic orbits. Each row corresponds to a distinct value of the parameter $\lambda$: from the first to the fifth row, $\lambda$ takes the values $0.01$, $0.03$, $0.05$, $0.07$, and $0.1$, respectively. Each column corresponds to a different parameter combination $\left(z, w, v\right)$, with specific values indicated in the upper left corner of each subgraph. Thus, the subgraphs represent the shape of the periodic orbits under fixed row parameters and the corresponding $\left(z, w, v\right)$ combinations. By comparing subgraphs within the same column across different rows, the effect of $\lambda$ on the periodic orbit can be examined; similarly, comparing subgraphs within the same row across different columns allows for the analysis of variations induced by changes in $\left(z, w, v\right)$. The most evident change is an increase in the number of ``zoom leaves'' $z$ in the orbit. The ``vertex number'' $v$ of the periodic orbit does not affect its shape.

\subsection{Gravitational wave radiation from periodic orbits}
We next investigate the gravitational radiation induced by the timelike particle moving along periodic orbits around the Schwarzschild-dS BH immersed in PFDM, effectively modeling an extreme-mass-ratio inspiral (EMRI) where the smaller body is much lighter than the central black hole. Consequently, the small object can be treated as a perturbation to the background spacetime, and its energy and angular momentum gradually change due to gravitational radiation. Since these changes remain small over several orbital periods, the adiabatic approximation is justified, allowing us to trace the periodic orbits and compute the corresponding gravitational waves. To this end, we adopt the kludge waveform method proposed in \cite{Tu2023, Babak2007} and evaluate the emitted gravitational waves up to quadratic order, following \cite{Maselli2022, Liang2023b}. This approach is particularly advantageous for analyzing gravitational waves in environments such as PFDM, where the orbiting mass dynamics near a supermassive black hole become relevant. Using the adiabatic approximation, the losses of energy and angular momentum are negligible over multiple periods, and we employ the Kuldge method to generate gravitational waveforms based on timelike geodesic motion \cite{Li2024}. This method incorporates both geodesic structure analysis and orbital periodicity to construct the waveform. Specifically, the adiabatically approximated values of energy and angular momentum for a given periodic orbit are obtained from the radial and azimuthal evolution derived from geodesic analysis, and the gravitational waveform is described via the quadrupole formula, expressed as
\begin{align}
h_{ij}=\frac{4\eta M}{D_{L}}\left(v_{i}v_{j}-\frac{m}{r}n_{i}n_{j}\right),\label{35}
\end{align}
where $M$ and $m$ are the masses of the supermassive BH and the timelike particle, respectively, and $D_{L}$ represents luminosity distance of the EMR system. In addition, $\eta=Mm/\left(M+m\right)^{2}$ is the symmetric mass ratio, while $v_{i}$ and $n_{i}$ denote the spatial velocity and the radial unit vector of the motion of the timelike particle, respectively. The gravitational wave is then projected onto the detector-adapted coordinate system, in which the two polarization states $h_{+}$ and $h_{\times}$ satisfy
\begin{align}
h_{+}&=-\frac{2\eta}{D_{L}}\frac{M^{2}}{r}\left(1+\cos^{2}\iota \right)\cos\left(2\phi+2\zeta\right),\label{36}\\
h_{\times}&=-\frac{4\eta}{D_{L}}\frac{M^{2}}{r}\cos\iota \sin\left(2\phi+2\zeta\right),\label{37}
\end{align}
where $\phi$, $\iota$, and $\zeta$ denote the phase angle, the inclination angle, and the longitude of periastron, respectively. 

\begin{figure}[H]
\centering{\includegraphics[scale=0.35]{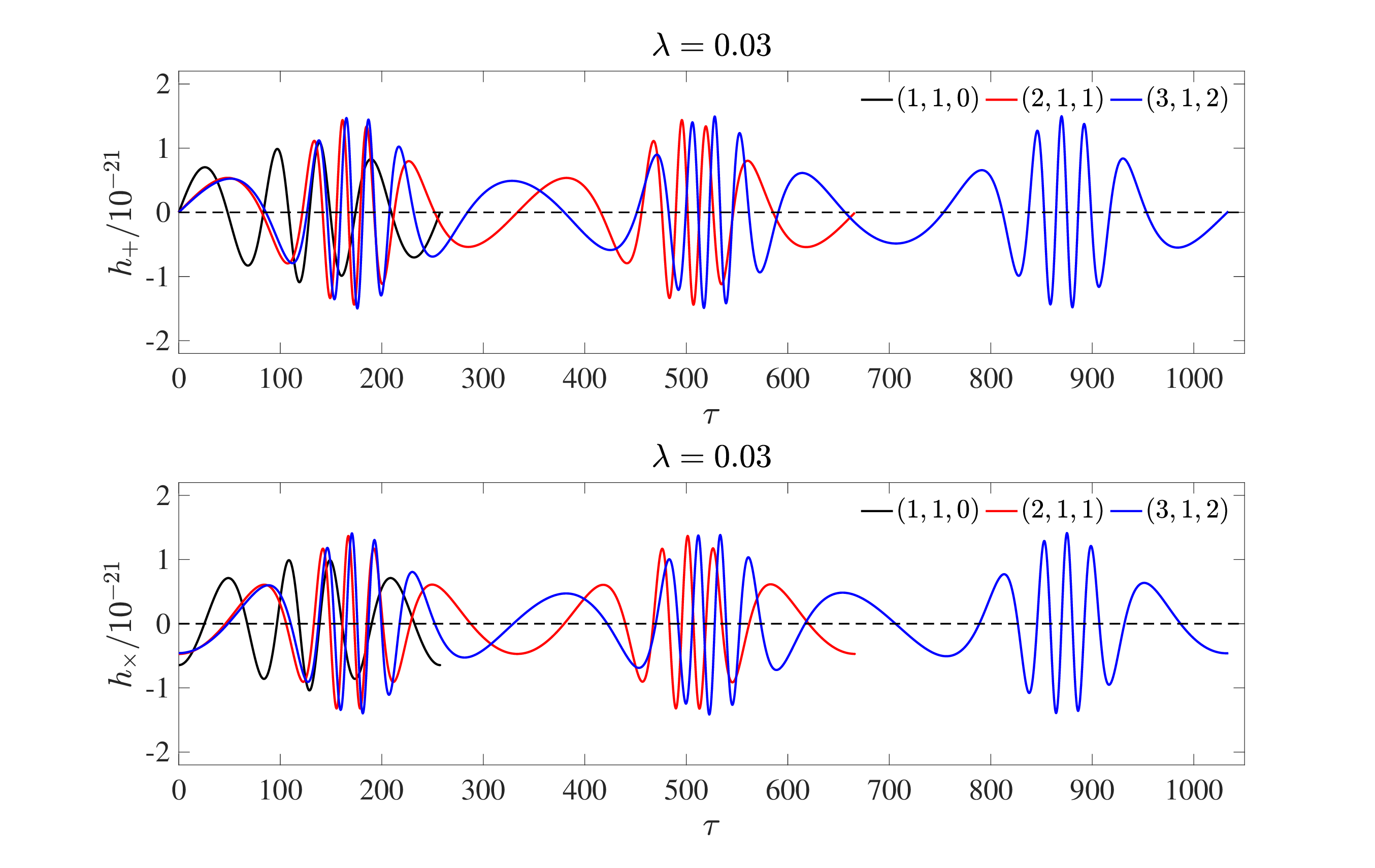}}
\caption{The gravitational wave of the Schwarzschild-dS BH immersed in PFDM with $\lambda=0.03$ and different values of  $(z, w, v)$. Specifically, the black line represents the gravitational wave with $q=1$; the red line represents the gravitational wave with $q=1+1/2$; the blue line represents the gravitational wave with $q=1+2/3$.}
\label{fig15}
\end{figure}

\begin{figure}[H]
\centering{\includegraphics[scale=0.35]{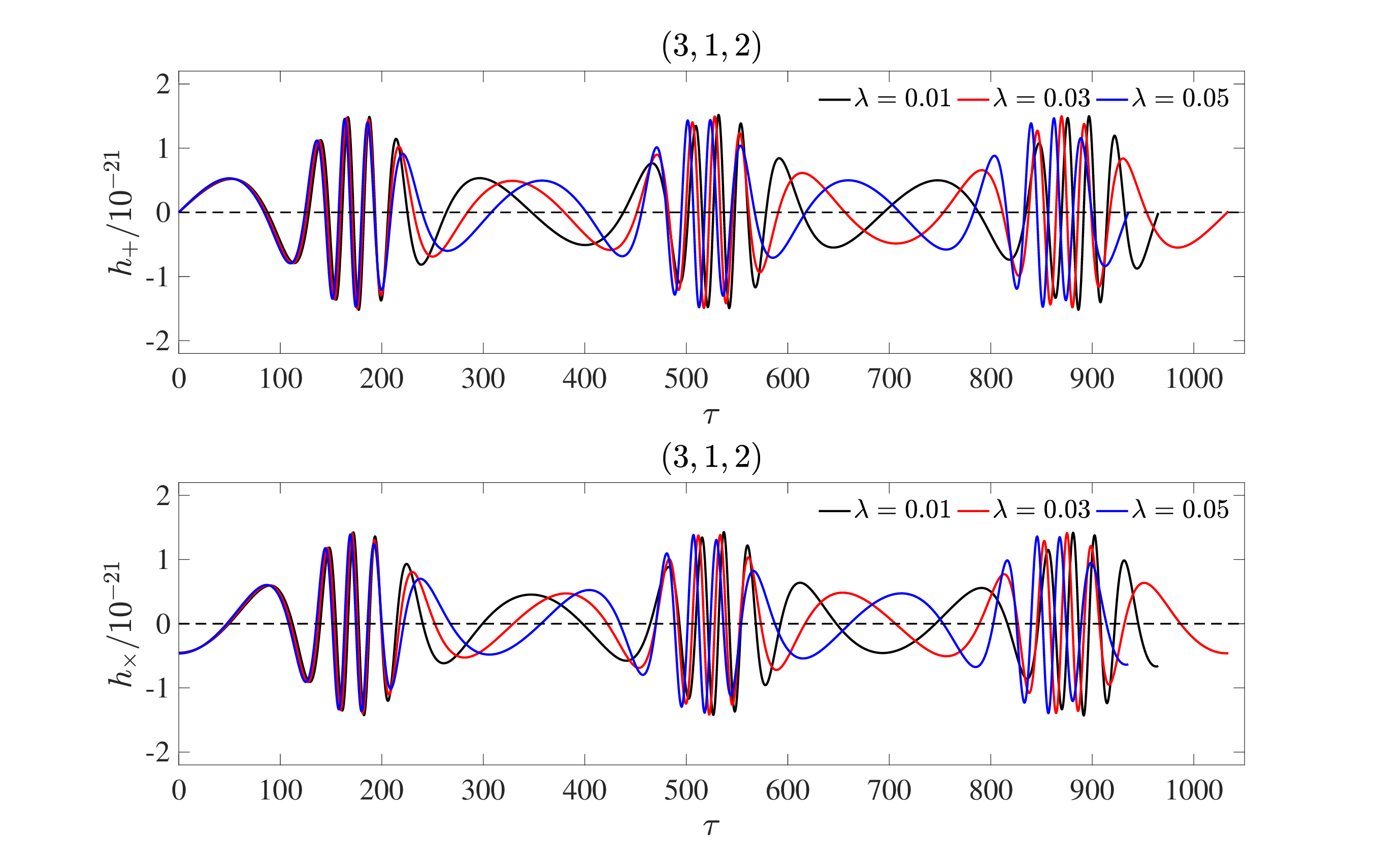}}
\caption{The gravitational wave of the Schwarzschild-dS BH immersed in PFDM with $(z, w, v)=(3, 1, 2)$ and different values of the parameter $\lambda$. Specifically, the black line represents the gravitational wave with $\lambda=0.01$ and $q=1+2/3$; the red line represents the gravitational wave with $\lambda=0.03$ and $q=1+2/3$; the blue line represents the gravitational wave with $\lambda=0.05$ and $q=1+2/3$.}
\label{fig16}
\end{figure}

To examine the gravitational waveforms of various periodic orbits and the impact of dark matter field on them, we take an EMR system made up of a supermassive Schwarzschild-dS BH immersed in PFDM $(M=10^{7}$ M$_{\odot})$ and a small component $(m=10$ M$_{\odot})$. Take $D_{L}=200$ Mpc and $\iota=\zeta=\pi/4$. Additionally, under the adiabatic approximation and neglecting radiation backreaction on periodic orbits, we restrict the calculation to a single complete orbital period to ensure precision. Figure \ref{fig15} presents the gravitational wave polarizations $h_{+}$ and $h_{\times}$ for periodic orbit samples with $E=0.96$ and $\lambda=0.1$. Over one complete period, alternating zoom and whirl phases appear in the gravitational waveforms, reflecting the zoom-whirl behavior characteristic of periodic orbits. Specifically, relatively quiet phases are exhibited by both $h_{+}$ and $h_{\times}$ during the highly elliptical zoom stage, with an intermediate phase following during the nearly circular whirl stage. As can be directly observed from each waveform, the number of highly elliptical phases corresponds precisely to the number of closed leaves in the orbit. Figure \ref{fig16} shows the gravitational waveforms for the fixed periodic orbit $(3,1,2)$ with different values of the parameter $\lambda$. These patterns indicate that within a certain range, increasing $\lambda$ lengthens the waveform period and suppresses the gravitational-wave amplitude. 

It is worth noting that our calculations, based on equation (\ref{35}) under the adiabatic approximation, do not yield precise gravitational waveforms, as they exclude multipole contributions beyond the quadratic order. Nonetheless, this method adequately captures the fundamental gravitational signals from periodic orbits. Early results show that their gravitational waves encode vital details of the whirl and zoom phases, offering a potential means to probe orbital characteristics. Moreover, this framework provides a clear visualization of realistic EMR systems, consisting of a stellar-mass object, a supermassive black hole, and surrounding the PFDM. Although the adopted approximation does not produce precise waveforms, it effectively captures the zoom-whirl characteristics of trajectories in gravitational wave signals. The adiabatic approximation proves reliable for single-period analysis, and the quadrupole formula used for wave polarization assumes low eccentricity, which necessarily omits higher-order terms required for more accurate modeling. Consequently, while this method works well for the physical description of gravitational waves over a single period, it fails over multiple periods, where higher-order and non-adiabatic effects become essential. As noted in \cite{Zelenka2020}, when the spin of the in spiraling compact object is considered in an EMRI system, prolonged resonances accompanied by chaotic motion may arise, producing gravitational waveforms with distinct resonance features. Thus, a general description of gravitational waves in EMR systems must incorporate higher-order and non-adiabatic terms, thereby enhancing our broader understanding of such systems. Future space-based gravitational wave observatories are expected to provide the observational data needed to test theoretical models and offer deeper insights into black hole properties and their connections to dark matter.

\section{Discussion and conclusions}\label{se-5}
This paper examines the geodesic motion of timelike particles in the Schwarzschild-dS BH immersed in PFDM, with emphasis on the general properties of bound orbits confined between the innermost and outermost stable circular orbits, as well as the gravitational-wave emission from periodic orbits.

Algebraic solutions and numerical analyses on the lapse function in the black hole spacetime metric show that, under certain conditions, the spacetime may possess two, one, or zero event horizons. The larger one corresponds to the cosmological horizon, while the smaller one is the event horizon. When the black hole mass reaches its maximum value, the two horizons coalesce into a single one. Furthermore, the analytical expressions of the Ricci scalar, the square of the Ricci tensor, and the Kretschmann scalar of this spacetime are provided to examine the gravitational effects of the black hole. The results indicate that both the cosmological constant and the PFDM parameter serve as significant source terms for the spacetime curvature, and an inevitable curvature singularity emerges at the radial coordinate $r = 0$.

To study the geodesic motion of timelike particles around BHs, the geodesic equation is derived from the Lagrangian via the Euler-Lagrange formalism, and the corresponding effective potential is obtained. Analysis of this potential shows abundant bound orbits: extrema of the potential correspond to stable and unstable circular orbits, while inflection points determine the innermost and outermost stable circular orbits. Numerically, the dependence of the radii, angular momentum, and energy of these orbits on the cosmological constant and the dark matter field is investigated. It is found that both the innermost and outermost stable circular orbit radii decrease with the dark matter parameter; the innermost radius increases with the cosmological constant, whereas the outermost radius decreases. The angular momenta of both orbits decrease with both parameters. The energies increase with the dark matter parameter but decrease with the cosmological constant. Additionally, the phase portraits are plotted to explore the long-term dynamical behavior of the system.

In particular, the bound orbit between the innermost and outermost stable circular orbits deserves focused consideration. There exists a mutual constraint between the energy and angular momentum of the particle motion. Accordingly, the allowed regions in the energy-angular momentum parameter space (i.e.,  $(E,\;L)$ region) were plotted for various values of the PFDM parameter. By applying the Levin orbit classification method and combining it with observational data on the S2 star orbiting Sagittarius A* at the Galactic center, the PFDM parameter is constrained to $0 < \lambda < 7.726412 \times 10^{5}$ M$_\odot$. These results indicate that neither the dark matter field nor the cosmological constant can be completely ruled out by realistic astrophysical environments.

Furthermore, the periodic orbits between the innermost and outermost stable circular orbits were analyzed and plotted. By simulating the EMRI system, the gravitational radiation of periodic orbits was investigated. The results indicate that as the zoom number increases, the orbital period exhibits an increasing trend, while the amplitude of the gravitational wave slightly increases. Moreover, the effect of the PFDM parameter on the orbital period is nonlinear; as this parameter increases, the gravitational wave amplitude shows a slight decrease, and the phase difference between waveforms remains relatively small. In view of the prevalent rotational states of supermassive black holes in astrophysical environments, the present model will be generalized to a more observationally relevant regime by including the effects of black hole spin. Further analyses will be directed toward the orbital dynamical evolution and the associated gravitational wave emissions, which are expected to offer new insights into the nature of dark matter and dark energy.

\section*{Acknowledgements}
This research was funded by the National Natural Science Foundation of China (NSFC) through grant nos. 12172322, the ``High-end Talent Support Program'' of Yangzhou University, and the Basic Research Program of Jiangsu, China (BK20250921).

{\small

\end{document}